# EXPONENTIAL DECAY OF CONCENTRATION VARIANCE DURING MAGMA MIXING: ROBUSTNESS OF A VOLCANIC CHRONOMETER AND IMPLICATIONS FOR THE HOMOGENIZATION OF CHEMICAL HETEROGENEITIES IN MAGMATIC SYSTEMS

Stefano Rossi[1*], Maurizio Petrelli[1], Daniele Morgavi[1], Diego González-García[1], Lennart A. Fischer[2], Francesco Vetere[1], Diego Perugini[1]

[1] *Department of Physics and Geology, University of Perugia, Piazza Università, 06100 Perugia, Italy*

[2] *Institute of Mineralogy, Leibniz Universität Hannover, Callinstrasse 3, 30167, Hannover, Germany*

\* *Corresponding Author*

Stefano Rossi

Tel. +39 0755852601

e-mail: stefano.rossi1@studenti.unipg.it

**Abstract**

The mixing of magmas is a fundamental process in the Earth system causing extreme compositional variations in igneous rocks. This process can develop with different intensities both in space and time, making the interpretation of compositional patterns in igneous rocks a petrological challenge. As a time-dependent process, magma mixing has been suggested to preserve information about the time elapsed between the injection of a new magma into sub-volcanic magma chambers and eruptions. This allowed the use of magma mixing as an additional volcanological tool to infer the mixing-to-eruption timescales. In spite of the potential of magma mixing processes to provide information about the timing of volcanic eruptions its statistical robustness is not yet established. This represents a prerequisite to apply reliably this conceptual model. Here, new chaotic magma mixing experiments were performed at different times using natural melts. The degree of reproducibility of experimental results was tested repeating one experiment at the same starting conditions and comparing the compositional variability. We further tested the robustness of the statistical analysis by randomly removing from the analysed dataset a progressively increasing number of samples. Results highlight the robustness of the method to derive empirical relationships linking the efficiency of chemical exchanges and mixing time. These empirical relationships remain valid by removing up to 80% of the analytical determinations. Experimental results were applied to constrain the homogenization time of chemical heterogeneities in natural magmatic system during mixing. The calculations show that, when the mixing dynamics generate millimetre thick filaments, homogenization timescales of the order of a few minutes are to be expected.

1. Introduction

Magma mixing is a major petrogenetic process concurring in the generation of the wide compositional diversity of igneous rocks on Earth (Bateman, 1995; Bacon, 1986; Eichelberger, 1978; Sparks et al., 1977) in both the plutonic and volcanic environment (Albert et al., 2015; Anderson, 1976, 1982; De Rosa et al., 1996; Kratzmann et al., 2009; Perugini and Poli, 2005; Wiebe, 1994). In this work, according to Perugini and Poli (2012) we refer to magma mixing as the process combining the physical dispersion of the two magmas and the development of the chemical exchanges between them.

The evidence of magma mixing processes remained recorded by a range of structural and textural features in igneous rocks, including the occurrence of mineral phases showing thermo-chemical disequilibria (Anderson, 1984; Didier and Barbarin, 1991; Hibbard, 1981; Wada, 1995; Wallace and Bergantz, 2002), magmatic enclaves dispersed into compositionally different host rocks (Bacon, 1986; Vetere et al., 2015), and bandings of different magma compositions (Flinders and Clemens, 1996; Morgavi et al., 2016).

Geochemically, magma mixing is witnessed by extreme compositional variations of major and trace elements and isotopes, which can occur in the rocks even at very short length scales, of the order of millimetres or micrometres (Montagna et al., 2015; Wiesmaier et al., 2015; Laeger et al., 2017). The presence of compositionally different domains at such short length scale can cause diffusive fractionation processes of chemical elements that, due to their different mobility in magmas (Perugini et al., 2006), generate volumes of melts whose compositional variation is difficult to reconcile with classical geochemical models (e.g. Fourcade and Allegre, 1981).

The use of numerical models and experimental petrology provided a powerful tool in the study of magma mixing, and several attempts have been made to capture the most relevant parameters involved during the interaction between magmas (Kouchi and Sunagawa, 1985; Laumonier et al., 2014a-b; 2015; Bergantz et al., 2015; Schleicher et al., 2016). These studies also highlighted an extreme complexity of the mixing process in space and time due to the interplay of the fluid dynamic regime and chemical exchanges between the interacting magmas.

The complexity of magma mixing processes has been shown to follow chaotic dynamics characterized by scale-invariant (fractal) compositional patterns (Flinders and Clemens, 1996; Perugini et al., 2003; Perugini and Poli, 2005; Petrelli et al., 2011; 2016). The chaotic nature of magma mixing processes depends on the kinematics governing the flow fields of the magmatic mass. In particular, the fundamental process producing physical mixing is the stretching and folding of two fluids. The stretching and folding process also represents the basic dynamics leading to chaotic behavior. As stretching and folding dynamics develop in time, an intricate lamellar pattern of flow structures is generated in which the interface area between magmas grows exponentially (e.g. Ottino, 1989). This is a pre-requisite for efficient mixing because chemical exchanges are enhanced through diffusion. Several physical processes can cause stretching and folding dynamics in magmas. The most important are: convective motions and plume-like dynamics of a light magma ascending into a denser one (e.g. Bateman, 1995; Cruden *et al.*, 1995; Snyder and Tait, 1996; Cardoso and Woods, 1999), forced and fountain-like dispersion (e.g. Campbell and Turner, 1986), and ascent of magmas towards the Earth surface in dikes and channels (e.g. Koyaguchi, 1985; Blake and Campbell, 1986; Koyaguchi and Blake, 1989). These processes have been studied using both experimental and numerical approaches highlighting that they are capable of generating stretching and folding dynamics between fluids (e.g. Metcalfe et al., 1995; Hydon, 1995; Meleshko and Van Heijst, 1995).

An important point emerging from magma mixing studies is that it is a time dependent process, i.e. the longer the mixing time, the more homogeneous the mixture. For this reason, it has recently been suggested, on the basis of high-temperature experiments performed with natural melts, that the compositional variability due to magma mixing might be utilized to define new geo-chronometers to estimate the mixing-to-eruption timescales (Perugini et al., 2010; 2015). However, in order to derive empirical relationships to constrain eruption timescales, a thorough assessment of the reliability of magma mixing experiments in terms of robustness and reproducibility of the results is needed. This is mostly due to the large experimental difficulties arising when dealing with such a complex process at high-temperature and with high-viscosity natural melts. These difficulties are further amplified by the chaotic nature of the mixing process that, by definition, might impede the reliable study of the long-term evolution of the process (e.g. Strogatz, 2001).

Motivated by these considerations, we performed a new set of chaotic magma mixing experiments using natural compositions with the aim to develop a robust statistical framework for the construction of mixing-to-eruption geo-chronometers. We evaluate the degree of reproducibility of mixing experiments, their statistical robustness and reliability. In particular, experiments at different mixing times and at condition relevant for natural magmatic systems were performed. The degree of reproducibility of experimental results was tested repeating one experiment using the same starting conditions and comparing the compositional variability of major elements. This allowed us to evaluate the robustness of the empirical relationships relating the degree of homogeneity of the samples against the mixing time. We further tested the robustness of our statistical analysis by randomly removing from the analysed experimental samples a progressively increasing number of samples and verifying the quality of the relationships obtained empirically. We then applied the obtained results to constrain the homogenization time of chemical heterogeneities in natural systems during the magma mixing process.

## 2. Materials and Methods

*2.1 Starting Materials*

Two natural lava samples from the island of Vulcano (Aeolian archipelago, Italy; (e.g. Keller, 1980; Gioncada et al., 2003) were used as end-members in the experiments. The least evolved (more mafic) end-member is a shoshonite sampled at the Vulcanello lava platform (hereafter shoshonitic composition SC; Davì et al., 2009). The most evolved is a high-K rhyolitic obsidian from the Pietre Cotte lava flow, belonging to La Fossa cone (hereafter rhyolitic composition RC; De Astis et al., 1997; Clocchiatti et al., 1994; Piochi et al., 2009). The chemical compositions of the shoshonitic and rhyolitic end-members are listed in Table 1.

Alteration-free rock samples were cleaned with distilled water and then crushed in an agate mortar to obtain a fine-grained powder. End-member powders were subsequently homogenized through two cycles of melting at 1600°C for 4 hours in a high temperature furnace (Nabertherm® HT 04/17) at ambient pressure using a $Pt_{80}Rh_{20}$ crucible followed by crushing and fine powdering in an agate mortar (e.g. Morgavi et al., 2015). End-member melts were characterized for their rheology using an Anton Paar RheolabQC viscometer installed at the Department of Physics and Geology of the University of Perugia (Italy). Viscosity of end-member melts at experimental temperature (1200 °C) are given in Table 1.

*2.2 Experimental Apparatus*

The experimental device used for chaotic mixing experiment, named Chaotic Magma Mixing Apparatus (COMMA; Morgavi et al., 2015) and hosted at the Department of Physics and Geology of the University of Perugia (Italy), was specifically designed to work with viscous (up to $10^8$ Pa s) natural and synthetic melts up to a temperature of 1500°C. The COMMA produces chaotic dynamics in the mixing system using a stirring protocol known in the literature as Journal Bearing System (JBS; e.g. Swanson and Ottino, 1990). Our high-temperature system consists of a $Pt_{80}Rh_{20}$ crucible (i.e. the outer cylinder), filled with the two end-member compositions and an inner, off-centred spindle (i.e. the inner cylinder; Fig. 1). Two parameters define the geometry of the COMMA assembly: (a) the ratio of the radii between the two cylinders, $r=R_{in}/R_{out}=1/3$, and (b) the eccentricity ratio to the outer cylinder $\varepsilon=\delta/R_{out}=0.3$, where $\delta$ is the distance between the centres of the inner and outer cylinders ($R_{in}$ and $R_{out}$) (Fig. 1). The alternate rotation of the crucible and the spindle in opposite directions triggers chaotic mixing within the system (e.g. Swanson and Ottino, 1990; Galaktionov et al., 2002). The time-periodic motion of the crucible and the spindle introduces two additional parameters: the rotation angle $\theta$ of the outer cylinder (i.e. the crucible) and the ratio $\Omega$ between the rotation angles of the outer and the inner cylinders (i.e. the crucible and the spindle). For the present work, $\theta$ and $\Omega$ were fixed to $2\pi$ and 3 respectively. The main fluid-dynamical parameters characterizing COMMA experiments are summarized in Table 2. The reader is redirected to the work of Morgavi et al. (2015) for a thorough description of the experimental apparatus.

Experiments were performed using volume percentages of RC and SC end-members of 88% and 12%, respectively. At completion of each experiment, the furnace holding the experimental sample was rapidly moved vertically while maintaining the crucible and the spindle at the original position. This allowed a very rapid quenching (cooling rate of ca. 50°C/min) of experimental melts, leading to the formation of a glassy experimental product. The quenched glasses were cored out from the outer cylinder, set in epoxy, cut into slices of ca. 4.0 mm thickness, and analysed by electron microprobe to study the space and time variability of chemical exchanges between the SC and RC end-member melts.

Mixing experiments were performed with different durations. Two experiments lasted 10.5 hours whereas the third experiment lasted 42 hours. In the following, we refer to these experiments as experiment A (10.5 hours), B (10.5 hours) and C (42 hours).

Our experiment started with two homogeneous contrasting multi-component silicate melts. With time, the initial elemental concentrations in the end-members are expected to drift toward the hybrid of this system. The estimation of the hybrid composition was obtained by the classical two end-member mixing equation (e.g. Langmuir et al. 1978):

$$C_H^i = C_A^i x + C_B^i (1-x) \qquad \text{Eq. 1}$$

where $C_H^i$, $C_A^i$ and $C_B^i$ are the concentrations of a given chemical element (*i*) in the hybrid, RC and SC melts, respectively, and *x* is the mass fraction of RC in the mixture. The concentration of the potential hybrid composition for each chemical element has been calculated by considering the end-member compositions reported in Table 1 and the initial mass fractions of the two melts (i.e. 88% of RC and 12% of SC).

*2.4 Analytical Methods*

Analyses were carried out on polished sections of each of the experimental samples. Elemental concentration profiles were obtained with a Cameca SX-100 electron microprobe at the Institute of Mineralogy of the Leibniz Universität Hannover (Germany). Operating conditions were an accelerating voltage of 15 kV, a beam current of 4 nA and a beam diameter of 10 μm, to minimize alkali loss in the glass analysis. Raw data were corrected with the software "Peak Sight" and "PAP" matrix (Pouchou and Pichoir, 1991). As standards, we used wollastonite for Si and Ca, $Al_2O_3$ for Al, $Fe_2O_3$ for Fe, MgO for Mg, rutile for Ti, albite for Na and orthoclase for K. The counting time was 10 s for each element. Precision and accuracy were determined by measuring VG-568 (rhyolite) and VG-2 (basalt) reference glasses. Obtained analytical errors are of the order of 4% for all the measured elements. A total of 24 profiles (11 for the experiment A, 7 for the experiment B and 6 for the experiment C) were obtained with a spacing between two successive analyses of 15 μm, for a total of 2134 analytical determinations.

*2.5 Normalized Variance*

A quantity commonly used in the fluid dynamics literature (e.g., Liu and Haller, 2004; Rothstein et al., 1999) to evaluate the degree of homogenization of fluid mixtures is the concentration variance $[\sigma^2(C^i)]$. The variance of concentration for a given chemical element ($\sigma^2(C^i)$) is given by:

$$\sigma^2(C^i) = \frac{\sum_{n=1}^{N}(C_n^i - \mu^i)}{N} \qquad \text{Eq. 2}$$

where $N$ is the number of the analytical determinations, $C_n^i$ is the concentration of element $i$ in the analytical determination $n$, and $\mu^i$ is the average composition for the element $i$. The concentration variance decreases with the increasing of mixing time, as the system progressively approaches homogeneity. As a consequence, the variance defined by Eq. 2 is a function of time ($t$). Note that $[\sigma^2(C^i)]$ also depends on absolute concentrations of each chemical element $i$. Given the different range of concentrations of elements in the experimental sample (Tab. 1), variance values have been normalized to the initial variance (e.g. Morgavi et al., 2013; Perugini et al., 2015). In the following, we refer to concentration variance, or simply variance, considering the following normalized quantity $[\sigma_n^2(C_i)]$:

$$\sigma_n^2(C^i) = \frac{\sigma^2(C^i)_t}{\sigma^2(C^i)_{t=0}} \qquad \text{Eq. 3}$$

where $\sigma^2(C_i)_t$ and $\sigma^2(C_i)_{t=0}$ is the variance of a chemical element ($C^i$) at the time $t$ and $t=0$ (i.e. before the beginning of the experiment), respectively.

*Concentration Variance Decay*

Following Perugini et al. (2015), the Concentration Variance Decay (CVD) for all chemical elements was modelled by the exponential:

$$\sigma_n^2(C_i) = C_0 \exp(-Rt) + C_1 \qquad \text{Eq. 4}$$

where $C_0$, $R$ and $C_1$ are fitting parameters and $t$ is the mixing time. In particular, $C_0$ and $C_1$ represent $\sigma_n^2(C_i)$ values at $t=0$ and $t=\infty$ respectively. Among the fitting parameters, $R$ is the most important for our purposes. It represents the rate at which concentration variance decays with time and it is a metric quantifying element mobility during mixing. In particular, it incorporates, in a single variable, the sum of a range of processes that affect element mobility, including: 1) partitioning of chemical elements into structurally different melts (e.g. Watson, 1976), 2) the dependence of diffusivities on multicomponent composition (e.g. Zhang, 2008), 3) the influence of advection on apparent diffusive fluxes (Perugini et al., 2006) and 4) the potential development of "uphill" diffusion patterns (Watson et al., 1984). Thus, the variable $R$ provides a first order cumulative assessment of element mobility.

*2.6 Data analysis*

Data analysis was performed using python (Oliphant, 2007) in the framework of Continuum Anaconda, a distribution specifically developed for scientific purposes (https://continuum.io). In detail, data importing and handling were performed using the Pandas Library (McKinney, 2012). CVD was modelled using the orthogonal distance regression algorithm (Brown and Fuller, 1990) implemented in the ScyPy package (van der Walt et al., 2011). Data were diagrammed using the Matplotlib (Hunter, 2007) and Seaborn libraries (Waskom et al., 2014).

**3. Results and Discussion**

Back-scattered electron images of representative portions of the experimental products are shown in Figure 2, displaying the action of chaotic dynamics by the presence of intricate mixing patterns represented by filaments alternating in composition. In particular, the images show the alternation of filaments with different thickness (from 800 to 5 µm) defining sub-portions of the mixing system in which the relative amount of interacting melts can be strongly variable. As an example, Figure 2A shows a large filament of the most mafic magma in which thin filaments of the most felsic one are dispersed. The opposite geometrical configuration, in the same experiment, can be seen in Figure 2B; here, a large filament of the most felsic melt is populated by thin filaments of the most mafic one. Also, the density of filaments is strongly variable within the same experiment. For example, Figure 2D shows the presence of four main filaments of the more mafic melt. In contrast, Figure 2C, belonging to the same experimental product, displays a much larger number of filaments (of the order of twenty). This is the natural consequence of the action of stretching and folding dynamics that characterize the space and time evolution of chaotic mixing processes (Ottino et al., 1988; Muzzio et al., 1992), as already observed in experiments and natural samples (e.g. Perugini et al., 2003; Perugini and Poli, 2012). Experimental samples were analysed for the compositional variation of major elements along transects crossing the filaments generated by the mixing process (Fig. 2). The complete analytical dataset containing all performed analyses is available as supplementary material (Table SM1). It consists of a total of 866, 780 and 488 chemical analyses for the experiment A (10.5 h), B (10.5 h) and C (42 h), respectively. The variability of selected elements (Si, Ca, Mg and K) along all analysed transects for experiments A, B and C is shown in Figure 3, where the end-member and the hybrid composition are also reported as a reference.

Qualitative observations of the chemical composition along the analysed transects in Figure 3 highlight that experiments A and B display similar patterns. In particular, it is notable that the composition of the RC end-member is still preserved in almost all filaments constituted by this component. On the contrary, element concentrations of the SC end-member disappeared in most of the filaments. As an example, $SiO_2$ contents are approximately 55 wt.% in the mafic filaments, slightly higher than the original composition (52.2 wt.%; Table 1). Experiment C shows a higher degree of homogenization with the concentration of all major elements shifting towards the hybrid composition.

Figure 4 displays representative binary diagrams for major elements ($K_2O$ vs. $SiO_2$, $TiO_2$ vs. $SiO_2$, $CaO$ vs. $SiO_2$, $MgO$ vs. $CaO$). Observation of these plots confirms the above considerations indicating that, for all element pairs of experiment A and B (performed at the same mixing time; $t$=10.5 h), data points display the same variability between the two end-members. On the contrary, results from experiment C (performed at longer mixing time; $t$=42 h) clearly exhibit a shrink in the compositional variability. In this case, the SC end-member is no longer detected in the mixing system (Fig. 4). In addition, most data points tend to converge towards the hybrid concentration (Fig. 3 and 4).

The degree of homogeneity, quantitatively evaluated using the normalized concentration variance [$\sigma_n^2(C_i)$], is reported as function of time (in hours) in Figure 5. According to the above definition, $\sigma_n^2(C_i)$ will be maximum at the beginning of the experiment and will decay to zero with time (Eq. 2). Figure 5 shows that the $\sigma_n^2(C_i)$ determined for the experiments A and B (10.5 hours) has the same value (within 2% of variability), with Na and Al showing minimal variations. The fact that $\sigma_n^2(C_i)$ values determined for the experiments A and B show no difference supports the good reproducibility of the experiments and corroborates the observations made for Figure 3 and 4.

The CVD was quantified for all analysed elements and the obtained $R$ values are reported in Figure 5. Evaluation of $R$ values indicates that the studied elements display ca. the same behaviour (0.151<$R$<0.175) in term of CVD (and, hence, mobility) with the only exception of Na, characterized by highest rates of decay ($R$=0.368). This is in agreement with theoretical and experimental results highlighting a higher mobility of Na with respect to the other major elements (Baker, 1990; Baker, 1991; Perugini et al., 2015).

In order to test the robustness of the method used to estimate $R$, a progressively increasing number of analytical determinations were randomly removed from the dataset and the values of $R$ further evaluated. This procedure was intended to provide information about the minimum number of samples to be analysed in order to obtain statistically significant results. In principle, with increasing number of removed samples, an increase of the 2-std value associated with the estimate of $R$ should be expected. Figure 6 shows that the increase of 2-std values remains below 18% for all the elements as the percentage of removed samples increases up to 80%. The only exception is Na, showing a faster increase of 2-std values.

To better understand the relation between 2-std values associated with the estimate of $R$ and the percentage of removed samples, Figure 7 shows the 2-std confidence intervals relative to the 80% (yellow areas) and 95% (orange areas) of removed analytical determinations. Figure 7 highlights that the 2-std confidence interval at 80% of removed samples is consistently close to the curve defined using the complete population of analyses. At 95% of removed samples, the 2-std confidence interval is obviously larger, but still defines a tight area for all elements, further highlighting the robustness of the method. Therefore, from the results presented above, the COMMA experimental device can be used to produce magma mixing experiments with a good level of accuracy and reproducibility. Experiments performed with the same starting conditions and durations show no significant difference in terms of concentration variance decay. Further, irrespective of the number of analytical determinations (in the limits highlighted above), the mixing system provides robust information upon the time development of the process. This consideration relies upon the fact that the development of chaotic dynamics produces fractal patterns in the mixing system. This has been documented in natural samples, numerical simulations and experimental studies (e.g. Perugini et al., 2003; 2006). The presence of fractal compositional domains implies the occurrence of the same (statistically speaking) patterns at many length scales. This guarantees that the sampling of such systems is always representative because the compositional modulation due to magma mixing is statistically the same at larger and smaller scales. This is clearly reflected by the results presented above where the values of *R* do not depend (considering the constraints defined above) upon the number of analysed samples. It has to be noted that similar results have been obtained studying natural samples (Perugini et al., 2015).

**4. Implications for natural systems**

The fact that the experimental system used here shows a good level of reproducibility of results represents a fundamental step, especially in the study of chaotic mixing processes, where the control of the fluid dynamics of the system is essential to guarantee accurate and reproducible results. In fact, in chaotic processes, very similar initial conditions can evolve in extremely different ways in short time. This is known as the "butterfly effect", typical of chaotic systems, in which nearby trajectories of the system evolution diverge exponentially with time (e.g. Strogatz, 2001). Due to this "sensitive dependence upon initial conditions" the design and development of experimental systems need to be accomplished with great precision. This is particularly true when dealing with high-temperature and high-viscosity natural melts that require challenging experimental approaches.

The fact that results presented above show a good reproducibility allows us to respond to a key question regarding the applicability of magma mixing experiments to estimate the mixing-to-eruption timescales for those magmatic systems in which magma mixing processes were frozen in time by the eruptions (e.g. Perugini et al., 2015). Indeed, we have shown that experiments are statistically robust and provide the opportunity to derive solid empirical relationships linking the extent of chemical exchanges and mixing time. This represents a significant achievement because these empirical relationships can be used to build new geochronometers to derive reliable mixing-to-eruption timescales.

An important achievement of our study is the estimation of the minimum amount of data required to obtain statistically significant results. We demonstrated that an extensive and detailed sampling is not a mandatory step in order to achieve robust estimations of the mixing-to-eruption timescales. This is based, however, on the assumption that the entire compositional variability due to magma mixing is analysed. As discussed above, the generation of fractal compositional patterns guarantees that the compositional variability is represented at many length scales. In principles, even the analysis of single thin sections should be able to preserve most of the compositional variability generated by magma mixing processes.

The performed magma mixing experiments can be also used to obtain new insights into the timescales required to homogenize the compositional heterogeneity due to magma mixing in magmatic systems. In the following we discuss the implications of our experiments on this topic using dimensional analysis.

In general, to achieve homogenization, an efficient action of chemical diffusion is a required condition. In the absence of advection, the homogenization time only depends on the dimension of the system and chemical diffusivities. In particular, a blob characterized by a dimension $L$ and a chemical diffusion coefficient $D$ will homogenize in a time $t_h$ proportional to $L^2/D$. We refer to $t_h$ as the homogenization time of the chemical heterogeneity.

In presence of advection, $t_h$ can be expressed as a function of the Reynolds ($Re$) and Schmidt ($Sc$) numbers (Raynal and Gence, 1997):

$$t_h \sim \frac{L^2}{\nu} \frac{1}{Re} f(ReSc) \qquad \text{Eq. 5}$$

where $\nu$ is the kinematic viscosity, $Re=VL/\nu$ and $Sc=\nu/D$.

In chaotic systems characterized by $ReSc>>1$, as in the case of most of natural magmatic systems, $t_h$ can be expressed as follow (Raynal and Gence, 1997):

$$t_h \sim \frac{L^2}{\nu} \frac{1}{Re} \ln(ReSc) \qquad \text{Eq. 6}$$

Equation 6 can be used to combine the results obtained experimentally with the dimension $L$ of the system and the strength of advective flows (through $Re$), i.e. the vigour of stretching and folding dynamics. In particular, considering filaments of different thickness of one magma dispersing into a host magma, it is possible to estimate the time required to attain homogenization as a function of $Re$, as mixing dynamics (i.e. the advective flows) develop in time. As here we are generalising our results, in the following we consider a representative range of $Re$ values ($10^{-9}$-$10^{-3}$) that can span the possible variability of $Re$ for natural systems. Figure 8 reports the homogenization times ($t_h$) in hours of filaments characterized by thicknesses of 0.001, 0.01, 0.1 and 1 m, as function of the Reynolds number ($Re$). Results indicate that a filament of 0.001 m (i.e.

one mm, comparable to the thickness of some filaments observed in the experimental samples) can be homogenised in ca. 10 minutes and 10 seconds at $Re=10^{-8}$ and $Re=10^{-6}$, respectively. As the thickness of filaments increases, also the homogenization time increases. As an example, the homogenization time for a 0.01 m thick filament is of the order of 18 h for $Re$ equal to $10^{-8}$. The same filament will be homogenized in ca. 1 seconds as $Re$ increases to $10^{-3}$. Ticker filaments of 0.1 and 1 m will homogenize in ca. 75 days and 21 years, respectively, at $Re = 10^{-8}$. Increasing $Re$ values to $10^{-3}$, will result in homogenization times of ca. 2 min and 3 h for a filament with a thickness of 0.1 and 1 m, respectively.

These results highlight that in the presence of chaotic conditions chemical heterogeneities can be readily erased in magmatic systems under appropriate fluid dynamic regimes. This has profound implications on our perception of the timescales of magma mixing processes. In fact, geologic processes generally appear as long-lasting events (of the order of thousands or millions of years). Here we have shown that magma mixing processes can develop in timescales that are orders of magnitude shorter than typical geological timescales.

An important implication emerging from the results presented above is that magma mixing can be characterized by the development of very efficient and rapid chemical exchanges, leading to the complete disappearance of the initial composition of the most mafic end-member (Fig. 3 and 4). Together with the production of large contact interfaces between the end-member magmas during chaotic mixing, additional factors may have played a role in such a quick disappearance of the mafic melt. Among those factors, the lower viscosity of the mafic melt is a crucial one. The mobility of chemical elements in low-viscosity mafic melts can be orders of magnitude larger than for high-viscosity felsic melts (e.g. Baker, 1990; Lesher, 1990; Zhang, 2008). This means that our ability to recover natural samples with the most mafic magma compositions mixed with a felsic magma may be seriously compromised. A possible approach to infer, at least partially, the composition of the end-members participating in the mixing process may come from detailed studies of compositional heterogeneity of crystals, providing that these crystals formed in the original end-member melts and that their composition was not altered during the progression of the mixing process. In this regard, Slaby et al. (2008; 2011) highlighted the importance of minerals in preserving information to reconstruct the composition of the end-member melts.

## 5. Conclusions

The results from the analysis of chaotic magma mixing experiments presented in this work highlight several crucial points:

i) The concentration variance of chemical elements decays exponentially with time leading to a rapid homogenization of compositional differences in the mixing system. This is due to the development of stretching and folding patterns during mixing that generate a scale-invariant (fractal) distribution of filaments with thicknesses down to very short length scales, of the order of tens of microns. The generation of large interfaces is the primary cause for the onset of efficient chemical exchanges between the two melts.

ii) The compositional variability generated by the mixing process is statistically robust. This is supported by the analysis of concentration variance that shows minimal variations by removing up to 80% of chemical analyses. This statistical robustness is due to the scale-invariant nature of mixing patterns and implies that the mixing system can be sampled and analysed at any length scale, providing statistically the same information.

iii) The experimental system (COMMA) used to generate magma mixing processes is suitable device that can be used to replicate with good approximation magma mixing processes between silicate melts. The fact that the mixing protocols can be tuned with precision makes this experimental apparatus an important tool to investigate the development of chaotic mixing processes, as observed in many natural outcrops (Flinders and Clemens, 1996; Perugini et al., 2003; Morgavi et al., 2016). The high level of precision of this machine is demonstrated by the good reproducibility of results obtained on repeated experiments, in terms of concentration variance of chemical elements.

iv) Given the good statistical robustness and repeatability of results from magma mixing experiments, the conceptual model of exponential decay of concentration variance with mixing time appears as a robust chronometer to be used in assessing the mixing-to-eruption time. In particular, results presented here provide a statistically robust framework to estimate the eruption timing for those eruptions in which magma mixing occurred between compositionally different melts without, or containing a few, crystals.

v) Starting from experimental results it is possible to generalize the relationship between the vigour of flow fields (i.e. the Reynolds number) and the time required to achieve homogenization during mixing. In particular, there is an inverse power-law relationship between the Reynolds number and homogenization time. In some cases, when the mixing dynamics generate millimetre thick filaments, it can be expected that homogenization timescales are very short, of the order of a few minutes. This implies that, if millimetre thick magma mixing structures are observed in natural rocks, the mixing process might have taken place shortly before the eruption. This appears in agreement with recent results highlighting that the mixing-to-eruption timescale can be very short, of the order of minutes (e.g. Perugini et al., 2015).

Although providing a first order assessment of the complexity associated with magma mixing processes the results presented this paper have some limitations that are worth discussing. Among them, one of the most important is that the experiments presented here were performed using silicate melts excluding the presence of minerals. In natural systems crystals of different minerals can be present, increasing the complexity of the process for several reasons. Minerals may occur in the end-member melts before the beginning of mixing or crystallization may start during mixing by the thermodynamic disequilibrium induced by this process. For example, the injection of a hot mafic magma into a felsic magma is likely to produce thermal and compositional disequilibria (Bateman, 1995; Sparks and Marshall, 1986). For example, minerals occurring in the lower temperature magma are expected to undergo a remelting/resorption process due to the increase in temperature. However, depending on the relative timing of resorption kinetics and attainment of new thermal equilibrium conditions, the resorption process may not be completed. Variably resorbed grains can act as nuclei for the growth of the same mineral species, but with a different composition, the latter depending on the initial composition of interacting magmas and their relative proportions. Compositional heterogeneity generated by the mixing process in the magmatic mass can also play a role in determining the extent of resorption and growth of minerals. Resorption and growth processes can be repeated in time leading to the formation of complex zoning patterns reflecting the intensity of chemical exchanges between the interacting magmas (e.g. Anderson, 1984; Hibbard, 1994; Perugini et al., 2003a; Perugini et al., 2005). As shown here, chemical exchanges depend of the fluid dynamic regime (i.e. the Reynolds number) characterising the mixing process. Minerals become, hence, witnesses of the fluid dynamics operating in the magmatic system during the development of magma mixing processes. Recent works highlighted the importance of detailed investigations of crystal compositional variability not only to reconstruct the fluid-dynamic regime governing the evolution of the igneous body, but also to understand the length-scale of the compositional variability induced by the mixing process (e.g. Perugini et al., 2003c; Pietranik and Koepke, 2009; Slaby et al., 2008; Slaby et al., 2011; Rocchi et al., 2016). An additional important role that can be played by minerals is that they might act as geometric perturbations of the flow fields, potentially enhancing the efficiency of the mixing process (Kouchi and Sunagawa, 1985; Laumonier, 2014a-b; 2015). This has been observed experimentally, where the mechanical distortion of flow fields tends to generate larger interface areas. As shown above, this is the prerequisite for enhancing chemical exchanges between compositionally different melts.

In the attempt to better understand also the effect of crystals during mixing, further chaotic mixing experiments will be performed using crystal-bearing end-members. This will tackle a new petrological challenge because the introduction of crystals will increase extraordinarily the complexity of the mixing system (e.g. Kouchi and Sunagawa, 1985; Laumonier, 2014a-b). In addition to the chemical exchange between melts, also the effect of potential dissolution and regrowth of minerals and their mechanical action in perturbing mixing flow fields in space and time will be considered.

A further limitation of the experimental approach presented above is that experiments can be performed, at present, only in dry conditions (i.e. excluding the role of volatiles dissolved in the melts). Although we believe that, due to the complexity of the processes overlapping during mixing, the best strategy is to proceed by adding degrees of freedom progressively, this represents another important aspect that should be considered in the future. Adding volatiles will be another challenging step and the most technically complex part will be probably represented by the construction of a new generation of experimental systems which are able to produce magma mixing processes under pressure in order to keep volatiles in solution in the silicate melts.

**Acknowledgements**

This research was funded by the European Research Council (ERC) Consolidator Grant ERC-2013-CoG No. 612776 – CHRONOS. F. Vetere wishes to acknowledge support from MIUR-DAAD Joint Mobility Project: (57262582). S. Rossi acknowledges the kind help received from J. Feige for the careful preparation of experiment sections at the Institute of Mineralogy in Hannover. Constructive comments by M. Laumonier, E. Slaby and G. W. Bergantz are gratefully acknowledged.

**Figure captions**

Figure 1: Representative schematic representation (not to scale) of the Chaotic Magma Mixing Apparatus (COMMA). A) Complete experimental setup showing the high-temperature furnace, the upper and lower motors for the rotation of the spindle and the crucible hosting the end-member melts, and the positioning of the experimental crucible; B) enlargement of the experimental crucible filled with the rhyolitic composition in which a glassy cylinder of shoshonitic composition is inserted. The off-centre positioning of the spindle is also shown; C) section of the crucible perpendicular to its vertical axis showing the relative position of end-member melts and the spindle. Geometrical parameters of the used experimental protocol are also shown [see text and Morgavi et al. (2015) for further details].

Figure 2: Representative segments of the experimental samples (shown as back-scattered

images) on which compositional transects were analysed. A-B) experiment A (10.5 h); C-D) experiment B (10.5 h); E-F) experiment C (42 h). The location of transects is reported as white lines showing the polarity of transects. Note in (C) and (D) the presence of bubbles. These are constituted by air that remained trapped while melting the natural rock compositions. However, as also shown by De Campos et al. (2011), they behave as passive markers and play no role in the timescale of the mixing experiments.

Figure 3: Compositional variation of some representative chemical element along the transects analysed for the three experiments performed at different times (experiment A, B and C). For the purpose of display, the several transects analysed for each experiment were concatenated to form a single series of compositional variation. Length of individual transects is reported on the horizontal axis for each chemical element. Black arrows indicate the polarity of the transects displayed in the corresponding panel in Figure 2. As a reference, the initial concentration of chemical elements in the end-member compositions and the theoretical hybrid composition are also shown.

Figure 4: Representative inter-elemental binary plots showing the compositional variability produced by the chaotic mixing process for the three experiments (experiment A, B, and C). As a reference, the initial concentration of chemical elements in the end-member compositions and the theoretical hybrid composition is also shown.

Figure 5: Variation of concentration variance as a function of mixing time for the analysed chemical elements fitted using Eq. 4 (see text). The value of *R* (i.e. the rate at which concentration variance decays with time) is reported for each element. Uncertainty on *R* is reported considering the uncertainty related to both the fitting and chemical analyses.

Figure 6: Plots showing the increase of 2-std values as a progressively larger number of samples from the experimental data set were removed.

Figure 7: 2-std confidence intervals relative to the removal of 80% (yellow areas) and 95% (orange areas) of analytical determinations.

Figure 8: Variation of the homogenization time (in hours) as a function of the Reynolds number for filaments of different thickness (L, in m) of a mafic melt dispersed into a felsic host melt.

**Table Captions**

Table 1: Concentrations (in weight %) of major elements in the rhyolitic and shoshonitic end-member glasses used in the chaotic mixing experiments. The viscosity of both end-members is also reported.

Table 2: Experimental parameters used for mixing experiments at T=1200°C. Values of Re: Reynolds Number; Sc: Schmidt Number; $R_{in}$: radius of the inner cylinder; $R_{out}$: radius of the outer cylinder; $V_{in}$: rotation velocity of the inner cylinder; $V_{out}$: rotation

velocity of the outer cylinder.

**Supplementary Material**

Table SM1: Complete analytical dataset containing all analytical determinations for Experiment A, B and C. A total of 2134 analytical determinations were obtained (866 for Experiment A, 780 for Experiment B and 488 for Experiment C).

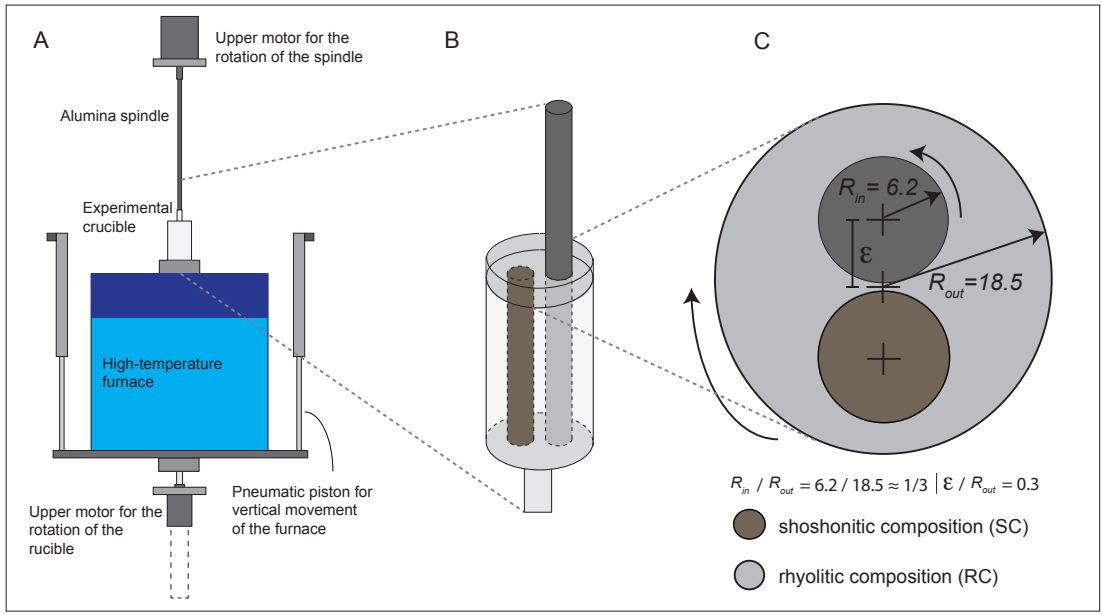

Figure 1

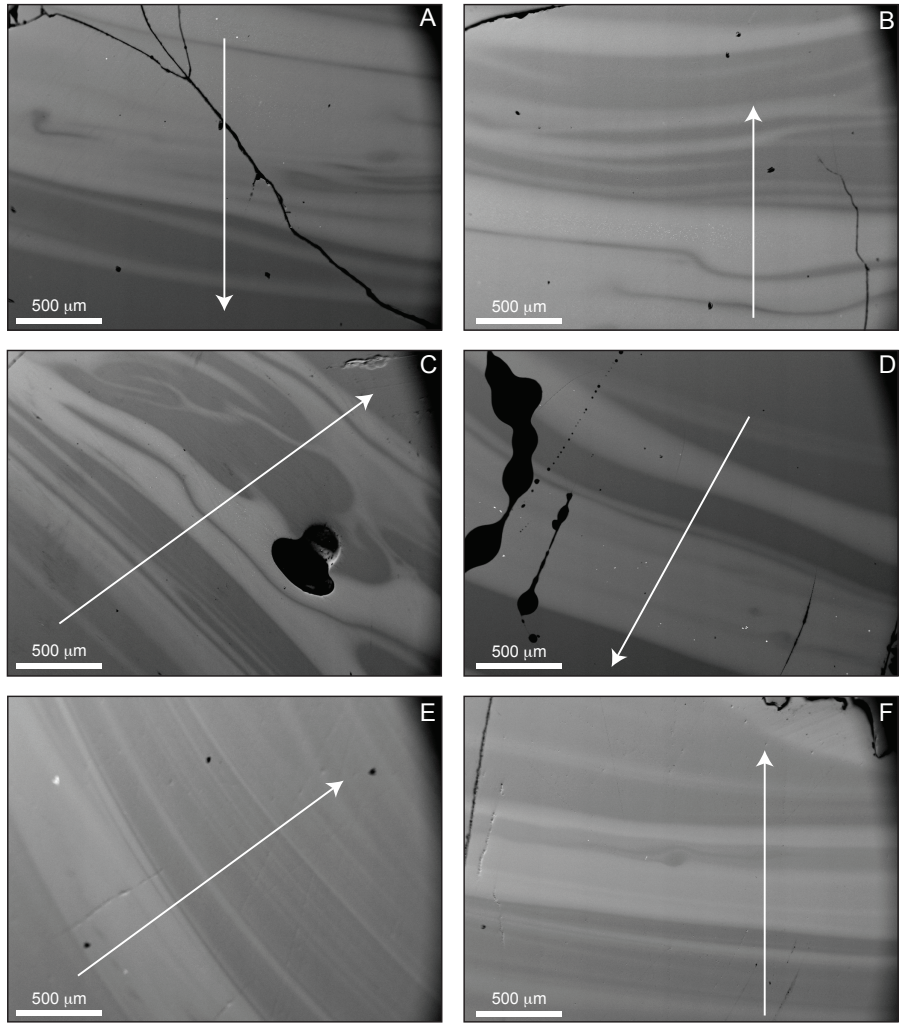

Figure 2

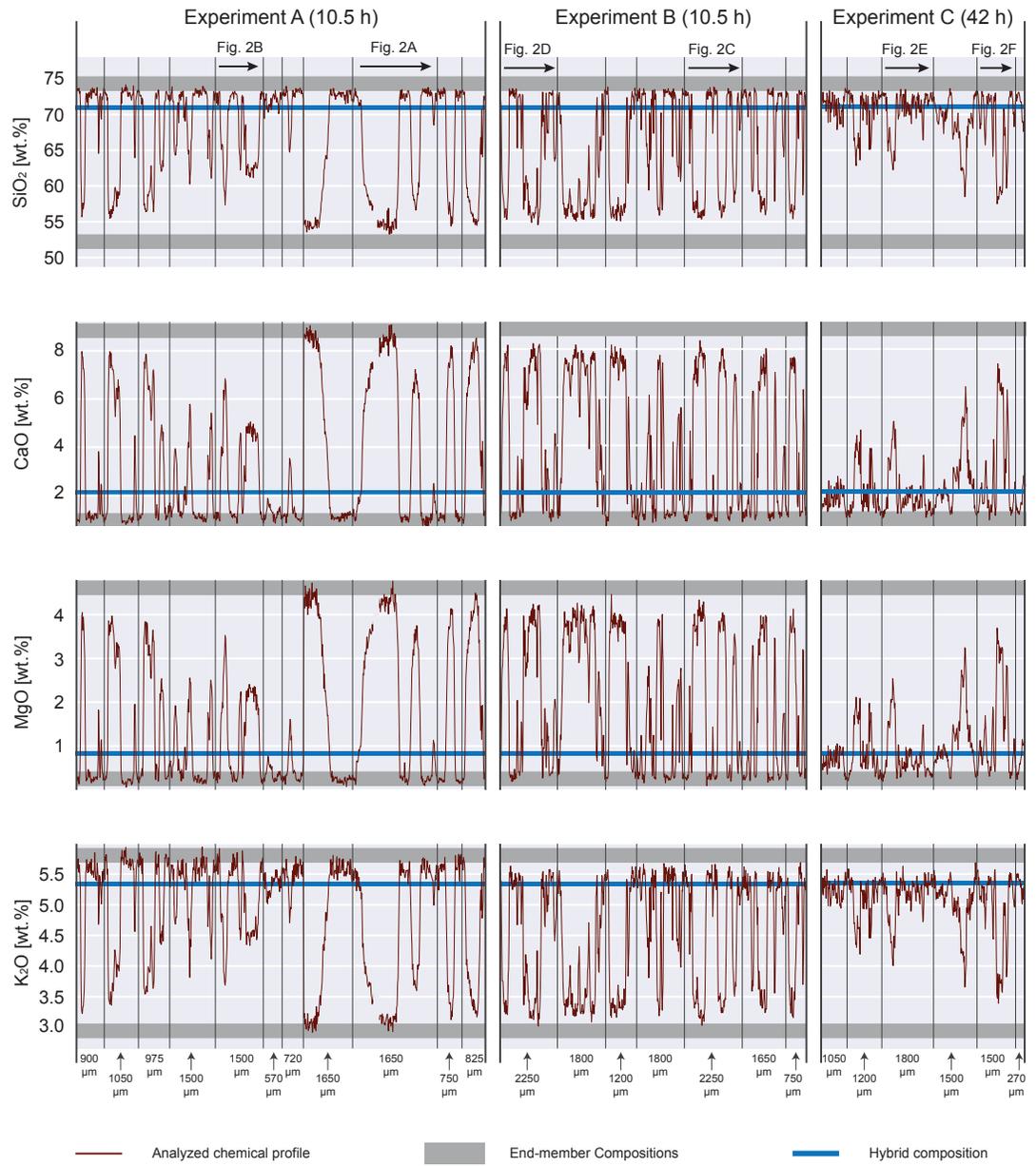

Figure 3

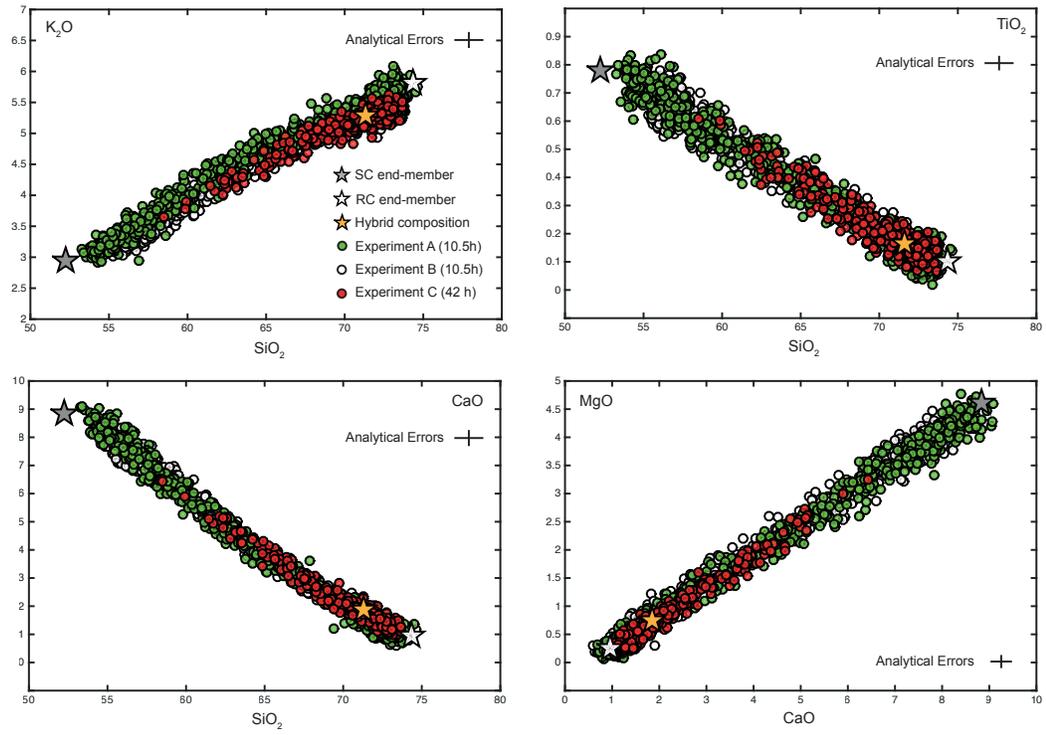

Figure 4

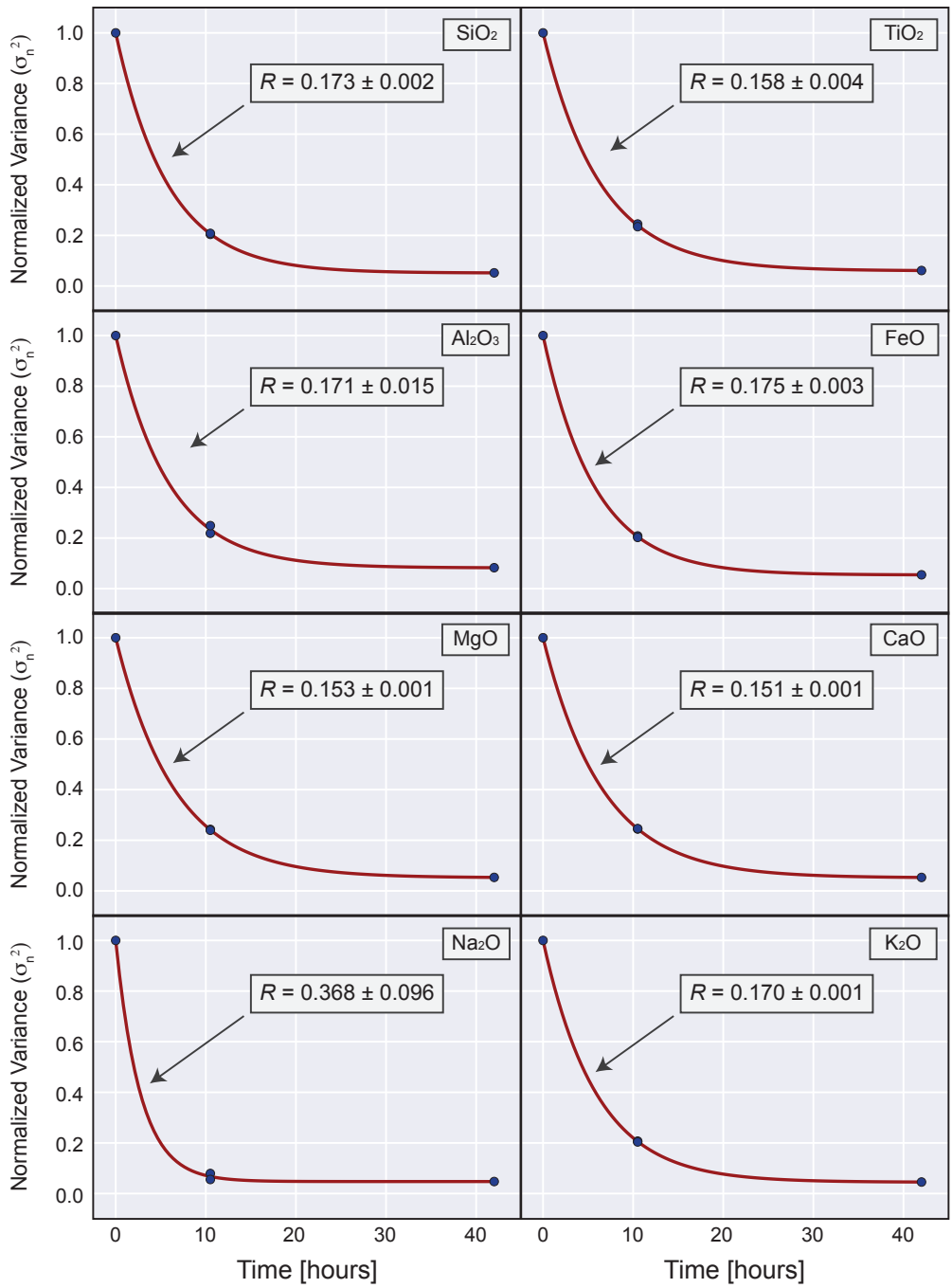

Figure 5

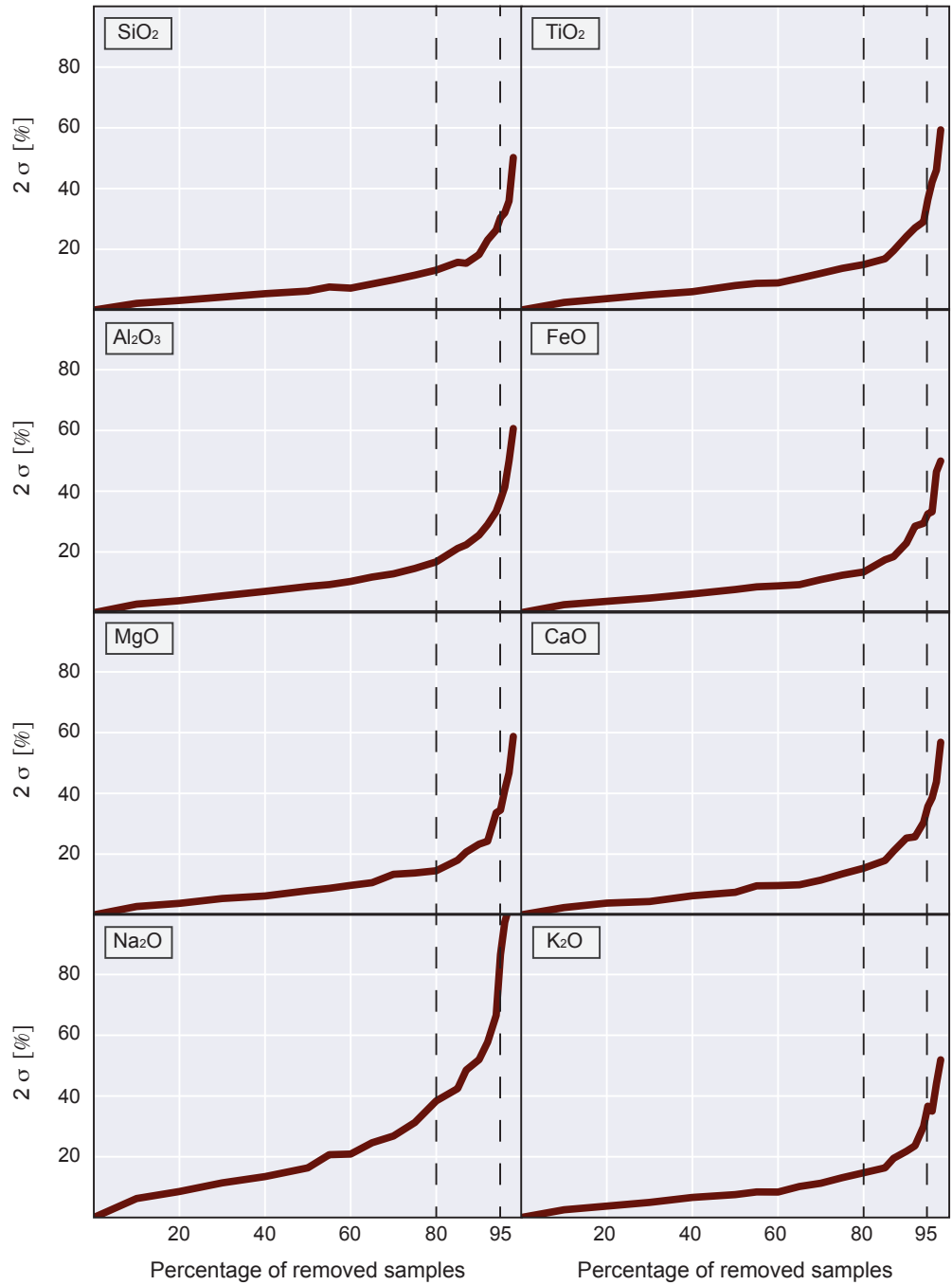

Figure 6

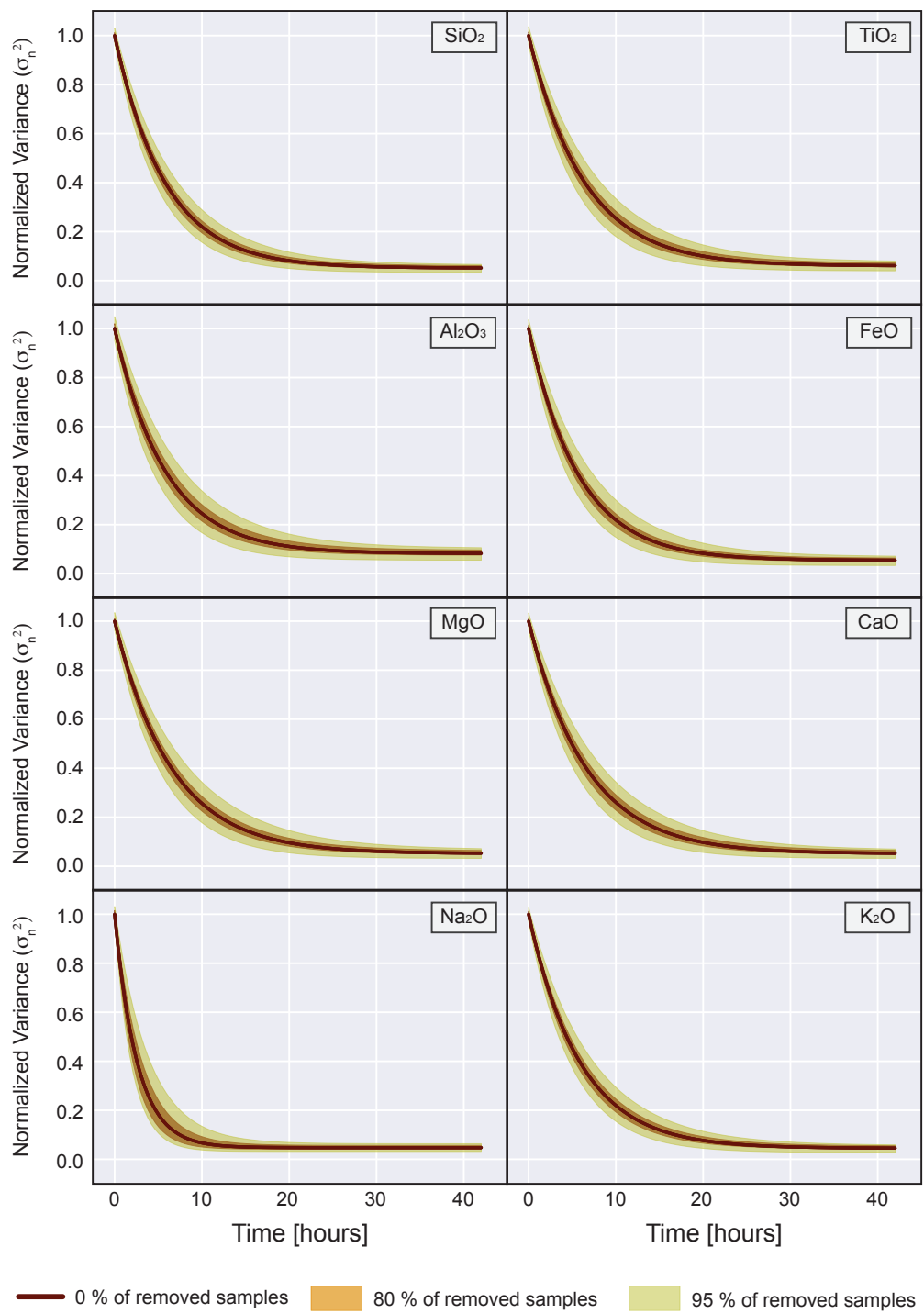

Figure 7

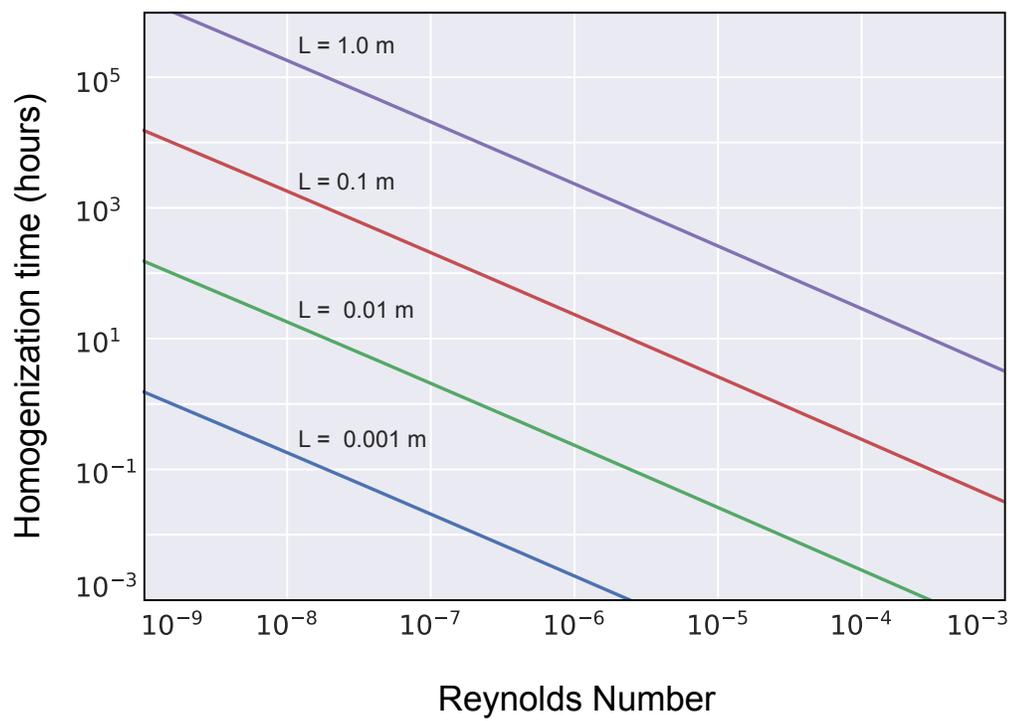

Figure 8

|  | Shoshonitic end-member (SC) wt% | Rhyolitic end-member (RC) wt% |
| --- | --- | --- |
| $SiO_2$ | 52.22 | 74.32 |
| $Al_2O_3$ | 16.38 | 13.21 |
| $K_2O$ | 2.94 | 5.81 |
| $TiO_2$ | 0.78 | 0.10 |
| $FeO_t$ | 8.84 | 1.60 |
| CaO | 8.84 | 0.96 |
| $Na_2O$ | 5.39 | 3.75 |
| MgO | 4.61 | 0.25 |
| **Total** | 100.00 | 100.00 |
| **Viscosity (Pa s) @ 1200 °C** | $1.8 \times 10^2$ | $1.6 \times 10^5$ |

Table 1

| Experimental Parameters @ 1200°C | |
|---|---|
| Re= $\rho VL/v$ | $6.3 \times 10^{-9}$ |
| Characteristic velocity V = $|V_{in}| + |V_{out}|$ (m/s) | $2.8 \times 10^{-5}$ |
| Characteristic Lenght L = $R_{out} - R_{in}$ (m) | $1.23 \times 10^{-2}$ |
| $R_{out}$ (mm) | 18.5 |
| $R_{in}$ (mm) | 6.2 |
| Diffusivity (m²/s) | $1 \times 10^{-13}$ |
| Schmidt (Sc) | $5.4 \times 10^{14}$ |

Table 2

Table SM1

| Point | SiO$_2$ | Al$_2$O$_3$ | K$_2$O | TiO$_2$ | FeO | CaO | Na$_2$O | MgO | Total |
|---|---|---|---|---|---|---|---|---|---|
| | | | | Profile 1 | | | | | |
| 1 | 71.96 | 13.62 | 5.31 | 0.15 | 2.55 | 1.59 | 4.17 | 0.33 | 100.00 |
| 2 | 72.95 | 13.79 | 5.42 | 0.14 | 1.81 | 1.30 | 4.19 | 0.39 | 100.00 |
| 3 | 72.62 | 13.61 | 5.33 | 0.17 | 2.22 | 1.39 | 4.22 | 0.44 | 100.00 |
| 4 | 71.34 | 13.58 | 5.16 | 0.17 | 3.30 | 1.47 | 4.34 | 0.57 | 100.00 |
| 5 | 71.88 | 13.60 | 5.20 | 0.14 | 2.64 | 1.58 | 4.34 | 0.59 | 100.00 |
| 6 | 72.12 | 13.26 | 5.44 | 0.14 | 2.95 | 1.51 | 4.00 | 0.45 | 100.00 |
| 7 | 73.13 | 13.92 | 5.33 | 0.11 | 1.35 | 1.45 | 4.14 | 0.45 | 100.00 |
| 8 | 71.64 | 13.48 | 5.16 | 0.17 | 2.52 | 1.66 | 4.56 | 0.78 | 100.00 |
| 9 | 71.78 | 13.39 | 5.43 | 0.21 | 2.48 | 1.82 | 4.18 | 0.52 | 100.00 |
| 10 | 71.99 | 13.65 | 5.28 | 0.19 | 2.21 | 1.81 | 4.27 | 0.61 | 100.00 |
| 11 | 71.81 | 13.46 | 5.47 | 0.12 | 2.58 | 1.56 | 4.36 | 0.54 | 100.00 |
| 12 | 71.81 | 14.24 | 5.31 | 0.15 | 2.13 | 1.51 | 4.20 | 0.64 | 100.00 |
| 13 | 71.95 | 13.53 | 5.33 | 0.17 | 2.39 | 1.57 | 4.25 | 0.56 | 100.00 |
| 14 | 70.47 | 14.12 | 5.11 | 0.24 | 2.85 | 2.14 | 4.29 | 0.78 | 100.00 |
| 15 | 71.45 | 14.00 | 5.35 | 0.16 | 1.99 | 1.92 | 4.46 | 0.67 | 100.00 |
| 16 | 73.02 | 13.46 | 5.57 | 0.11 | 1.85 | 1.25 | 4.27 | 0.42 | 100.00 |
| 17 | 72.58 | 13.72 | 5.40 | 0.09 | 2.21 | 1.42 | 4.23 | 0.35 | 100.00 |
| 18 | 72.55 | 13.59 | 5.32 | 0.12 | 2.02 | 1.31 | 4.57 | 0.53 | 100.00 |
| 19 | 68.90 | 14.18 | 4.98 | 0.30 | 3.53 | 2.53 | 4.47 | 1.01 | 100.00 |
| 20 | 71.97 | 14.26 | 5.28 | 0.12 | 1.75 | 1.51 | 4.40 | 0.64 | 100.00 |
| 21 | 71.63 | 13.76 | 5.26 | 0.19 | 2.43 | 1.43 | 4.48 | 0.55 | 100.00 |
| 22 | 73.33 | 13.13 | 5.39 | 0.11 | 1.75 | 1.57 | 4.24 | 0.49 | 100.00 |
| 23 | 72.38 | 13.46 | 5.29 | 0.18 | 2.09 | 1.47 | 4.36 | 0.61 | 100.00 |
| 24 | 68.97 | 14.53 | 4.97 | 0.29 | 3.01 | 2.32 | 4.75 | 0.92 | 100.00 |
| 25 | 71.71 | 13.49 | 5.00 | 0.18 | 2.84 | 1.83 | 4.28 | 0.67 | 100.00 |
| 26 | 70.33 | 14.44 | 5.10 | 0.22 | 2.63 | 2.13 | 4.10 | 0.86 | 100.00 |
| 27 | 70.69 | 14.00 | 5.13 | 0.15 | 3.17 | 1.80 | 4.30 | 0.76 | 100.00 |
| 28 | 71.49 | 13.73 | 5.26 | 0.17 | 2.02 | 1.87 | 4.56 | 0.62 | 100.00 |
| 29 | 73.12 | 13.49 | 5.34 | 0.14 | 2.00 | 1.51 | 3.83 | 0.57 | 100.00 |
| 30 | 72.86 | 13.31 | 5.25 | 0.10 | 2.05 | 1.44 | 4.44 | 0.53 | 100.00 |
| 31 | 71.57 | 13.79 | 5.34 | 0.17 | 2.74 | 1.71 | 3.87 | 0.77 | 100.00 |
| 32 | 71.27 | 13.67 | 5.15 | 0.24 | 2.26 | 2.28 | 4.16 | 0.77 | 100.00 |
| 33 | 69.68 | 13.76 | 5.23 | 0.18 | 3.62 | 2.21 | 4.40 | 0.91 | 100.00 |
| 34 | 72.74 | 13.34 | 5.27 | 0.12 | 1.86 | 1.73 | 4.29 | 0.65 | 100.00 |
| 35 | 72.21 | 13.23 | 5.42 | 0.14 | 2.63 | 1.52 | 4.32 | 0.53 | 100.00 |
| 36 | 71.58 | 13.55 | 5.25 | 0.22 | 2.47 | 1.61 | 4.32 | 0.66 | 100.00 |
| 37 | 69.85 | 14.17 | 5.11 | 0.20 | 2.84 | 2.45 | 4.23 | 1.00 | 100.00 |
| 38 | 71.22 | 13.80 | 5.11 | 0.17 | 2.88 | 1.76 | 4.20 | 0.70 | 100.00 |
| 39 | 72.16 | 13.79 | 5.19 | 0.19 | 1.99 | 1.76 | 4.30 | 0.62 | 100.00 |
| 40 | 67.86 | 15.18 | 5.02 | 0.28 | 3.18 | 2.72 | 4.62 | 1.05 | 100.00 |
| 41 | 68.53 | 14.51 | 5.05 | 0.33 | 3.30 | 2.43 | 4.50 | 1.03 | 100.00 |
| 42 | 69.46 | 14.57 | 5.13 | 0.30 | 2.82 | 2.13 | 4.51 | 0.96 | 100.00 |
| 43 | 72.30 | 13.53 | 5.47 | 0.13 | 2.42 | 1.42 | 4.03 | 0.48 | 100.00 |
| 44 | 71.44 | 13.92 | 5.40 | 0.24 | 2.33 | 1.72 | 4.32 | 0.63 | 100.00 |
| 45 | 71.59 | 13.58 | 5.22 | 0.17 | 3.03 | 1.56 | 4.21 | 0.50 | 100.00 |
| 46 | 69.18 | 14.56 | 5.05 | 0.23 | 3.13 | 2.46 | 4.13 | 1.00 | 100.00 |
| 47 | 68.64 | 14.50 | 4.98 | 0.26 | 3.55 | 2.59 | 4.37 | 0.98 | 100.00 |
| 48 | 69.66 | 14.02 | 5.02 | 0.18 | 3.55 | 2.09 | 4.40 | 0.98 | 100.00 |
| 49 | 68.68 | 13.88 | 4.94 | 0.22 | 4.21 | 2.37 | 4.53 | 0.90 | 100.00 |
| 50 | 70.59 | 13.68 | 5.15 | 0.20 | 2.97 | 2.22 | 4.16 | 0.80 | 100.00 |
| 51 | 70.49 | 13.64 | 5.24 | 0.24 | 3.32 | 1.88 | 4.30 | 0.82 | 100.00 |
| 52 | 70.53 | 14.09 | 5.05 | 0.12 | 2.94 | 2.07 | 4.40 | 0.80 | 100.00 |
| 53 | 71.19 | 13.15 | 5.10 | 0.10 | 3.16 | 1.86 | 4.38 | 0.83 | 100.00 |
| 54 | 68.24 | 14.71 | 5.09 | 0.25 | 3.51 | 2.34 | 4.64 | 0.98 | 100.00 |
| 55 | 71.29 | 13.64 | 5.19 | 0.18 | 2.68 | 1.75 | 4.40 | 0.71 | 100.00 |
| 56 | 72.54 | 13.77 | 5.38 | 0.18 | 2.06 | 1.53 | 3.96 | 0.54 | 100.00 |
| 57 | 71.84 | 13.92 | 5.33 | 0.13 | 2.28 | 1.43 | 4.22 | 0.48 | 100.00 |
| 58 | 73.03 | 13.69 | 5.31 | 0.15 | 2.03 | 1.34 | 4.13 | 0.31 | 100.00 |
| 59 | 73.69 | 13.44 | 5.53 | 0.06 | 1.86 | 1.17 | 3.87 | 0.38 | 100.00 |
| 60 | 73.50 | 13.20 | 5.44 | 0.16 | 1.68 | 1.01 | 4.68 | 0.23 | 100.00 |
| 61 | 73.36 | 13.73 | 5.40 | 0.09 | 1.70 | 1.24 | 3.91 | 0.24 | 100.00 |
| 62 | 73.06 | 13.69 | 5.33 | 0.09 | 1.86 | 1.17 | 4.26 | 0.30 | 100.00 |
| 63 | 73.27 | 13.80 | 5.53 | 0.12 | 1.46 | 1.10 | 4.10 | 0.26 | 100.00 |
| 64 | 73.61 | 13.52 | 5.45 | 0.08 | 1.44 | 1.09 | 4.23 | 0.26 | 100.00 |
| 65 | 73.01 | 13.77 | 5.46 | 0.12 | 1.85 | 1.31 | 4.21 | 0.27 | 100.00 |
| 66 | 72.98 | 13.71 | 5.44 | 0.18 | 1.99 | 1.19 | 4.21 | 0.21 | 100.00 |
| 67 | 73.26 | 13.82 | 5.59 | 0.11 | 1.75 | 1.23 | 3.89 | 0.26 | 100.00 |
| 68 | 72.81 | 13.25 | 5.47 | 0.12 | 2.67 | 1.01 | 4.28 | 0.30 | 100.00 |
| 69 | 73.39 | 13.46 | 5.27 | 0.12 | 1.96 | 1.29 | 4.12 | 0.39 | 100.00 |
| **Profile 2** | | | | | | | | | |
| 1 | 70.96 | 14.01 | 5.25 | 0.16 | 3.34 | 1.54 | 4.41 | 0.34 | 100.00 |
| 2 | 72.17 | 14.30 | 5.37 | 0.13 | 1.92 | 1.33 | 4.09 | 0.44 | 100.00 |
| 3 | 72.29 | 13.68 | 5.41 | 0.13 | 2.81 | 1.17 | 3.96 | 0.41 | 100.00 |
| 4 | 73.00 | 13.59 | 5.19 | 0.16 | 2.14 | 1.34 | 4.06 | 0.42 | 100.00 |
| 5 | 72.16 | 13.93 | 5.45 | 0.11 | 2.16 | 1.40 | 4.16 | 0.48 | 100.00 |
| 6 | 71.66 | 13.70 | 5.33 | 0.18 | 2.39 | 1.67 | 4.31 | 0.61 | 100.00 |
| 7 | 72.75 | 13.20 | 5.38 | 0.19 | 2.40 | 1.40 | 4.09 | 0.61 | 100.00 |
| 8 | 72.49 | 13.70 | 5.40 | 0.19 | 1.82 | 1.53 | 4.19 | 0.68 | 100.00 |

| | | | | | | | | |
|---|---|---|---|---|---|---|---|---|
| 9  | 71.53 | 13.19 | 5.21 | 0.22 | 2.80 | 2.06 | 4.22 | 0.78 | 100.00 |
| 10 | 68.29 | 14.41 | 4.91 | 0.32 | 3.02 | 3.02 | 4.36 | 1.39 | 100.00 |
| 11 | 66.37 | 14.38 | 4.69 | 0.34 | 4.08 | 3.74 | 4.63 | 1.70 | 100.00 |
| 12 | 66.71 | 15.15 | 4.70 | 0.38 | 3.62 | 3.49 | 4.40 | 1.49 | 100.00 |
| 13 | 65.65 | 15.20 | 4.58 | 0.39 | 3.70 | 4.07 | 4.46 | 1.77 | 100.00 |
| 14 | 65.21 | 15.14 | 4.52 | 0.37 | 4.37 | 3.78 | 4.49 | 1.86 | 100.00 |
| 15 | 63.48 | 15.31 | 4.36 | 0.44 | 4.84 | 4.24 | 5.21 | 1.91 | 100.00 |
| 16 | 62.83 | 15.68 | 4.38 | 0.39 | 5.04 | 4.42 | 5.03 | 2.11 | 100.00 |
| 17 | 64.97 | 15.31 | 4.30 | 0.40 | 4.33 | 3.95 | 4.69 | 1.90 | 100.00 |
| 18 | 65.38 | 14.85 | 4.49 | 0.44 | 4.72 | 3.79 | 4.63 | 1.70 | 100.00 |
| 19 | 65.71 | 14.63 | 4.82 | 0.32 | 4.34 | 3.73 | 4.58 | 1.62 | 100.00 |
| 20 | 65.68 | 15.04 | 4.63 | 0.28 | 4.18 | 3.68 | 4.79 | 1.69 | 100.00 |
| 21 | 67.41 | 14.51 | 4.61 | 0.32 | 4.18 | 3.20 | 4.28 | 1.51 | 100.00 |
| 22 | 65.27 | 14.80 | 4.61 | 0.37 | 4.38 | 4.14 | 4.56 | 1.77 | 100.00 |
| 23 | 65.86 | 14.74 | 4.63 | 0.43 | 4.17 | 3.82 | 4.52 | 1.63 | 100.00 |
| 24 | 66.74 | 14.72 | 4.62 | 0.25 | 4.24 | 2.84 | 4.74 | 1.60 | 100.00 |
| 25 | 63.62 | 15.46 | 4.19 | 0.42 | 4.63 | 4.64 | 4.64 | 2.09 | 100.00 |
| 26 | 72.21 | 13.78 | 5.17 | 0.15 | 1.63 | 1.98 | 4.28 | 0.67 | 100.00 |
| 27 | 72.62 | 13.69 | 5.38 | 0.13 | 1.98 | 1.56 | 4.11 | 0.53 | 100.00 |
| 28 | 71.38 | 13.90 | 5.22 | 0.21 | 2.52 | 1.71 | 4.47 | 0.60 | 100.00 |
| 29 | 72.16 | 13.87 | 5.24 | 0.13 | 2.19 | 1.53 | 4.34 | 0.53 | 100.00 |
| 30 | 73.12 | 13.28 | 5.23 | 0.17 | 2.06 | 1.26 | 4.36 | 0.48 | 100.00 |
| 31 | 72.92 | 13.70 | 5.13 | 0.14 | 2.01 | 1.49 | 3.91 | 0.58 | 100.00 |
| 32 | 69.17 | 14.39 | 4.94 | 0.28 | 2.80 | 2.41 | 4.62 | 1.12 | 100.00 |
| 33 | 72.22 | 13.75 | 5.24 | 0.16 | 2.12 | 1.44 | 4.20 | 0.63 | 100.00 |
| 34 | 71.74 | 13.57 | 4.93 | 0.20 | 2.24 | 2.02 | 4.66 | 0.63 | 100.00 |
| 35 | 71.27 | 14.03 | 5.24 | 0.21 | 2.41 | 1.86 | 4.17 | 0.66 | 100.00 |
| 36 | 71.70 | 13.40 | 5.18 | 0.20 | 2.55 | 1.65 | 4.52 | 0.77 | 100.00 |
| 37 | 70.66 | 14.23 | 5.04 | 0.18 | 2.41 | 2.09 | 4.41 | 0.79 | 100.00 |
| 38 | 71.19 | 14.10 | 5.09 | 0.19 | 2.81 | 1.78 | 4.26 | 0.48 | 100.00 |
| 39 | 72.76 | 13.23 | 5.36 | 0.14 | 2.18 | 1.44 | 4.22 | 0.67 | 100.00 |
| 40 | 70.06 | 13.76 | 4.93 | 0.20 | 3.37 | 2.11 | 4.43 | 0.80 | 100.00 |
| 41 | 68.59 | 14.17 | 5.08 | 0.27 | 3.68 | 2.39 | 4.65 | 1.04 | 100.00 |
| 42 | 70.62 | 14.17 | 5.11 | 0.17 | 2.75 | 1.86 | 4.48 | 0.81 | 100.00 |
| 43 | 72.59 | 13.48 | 5.37 | 0.09 | 2.35 | 1.26 | 4.38 | 0.47 | 100.00 |
| 44 | 72.55 | 13.47 | 5.41 | 0.13 | 2.14 | 1.47 | 4.13 | 0.55 | 100.00 |
| 45 | 71.16 | 13.82 | 5.15 | 0.15 | 2.97 | 1.69 | 4.38 | 0.63 | 100.00 |
| 46 | 66.72 | 14.66 | 4.69 | 0.31 | 3.47 | 3.65 | 4.61 | 1.67 | 100.00 |
| 47 | 65.09 | 15.04 | 4.46 | 0.42 | 4.18 | 3.88 | 4.58 | 1.98 | 100.00 |
| 48 | 70.54 | 13.90 | 4.83 | 0.20 | 2.55 | 2.20 | 4.67 | 1.10 | 100.00 |
| 49 | 66.25 | 14.92 | 4.94 | 0.38 | 3.82 | 3.24 | 4.62 | 1.52 | 100.00 |

| 50 | 65.19 | 14.80 | 4.46 | 0.34 | 4.82 | 3.88 | 4.78 | 1.54 | 100.00 |
|---|---|---|---|---|---|---|---|---|---|
| 51 | 65.96 | 14.81 | 4.54 | 0.42 | 4.28 | 3.69 | 4.35 | 1.77 | 100.00 |
| 52 | 67.62 | 14.30 | 4.69 | 0.27 | 4.36 | 3.16 | 4.17 | 1.33 | 100.00 |
| 53 | 68.86 | 14.24 | 5.07 | 0.25 | 3.27 | 2.53 | 4.67 | 1.07 | 100.00 |
| 54 | 72.47 | 13.37 | 5.13 | 0.14 | 1.87 | 1.59 | 4.61 | 0.72 | 100.00 |
| 55 | 71.35 | 13.69 | 5.35 | 0.19 | 2.38 | 1.53 | 4.66 | 0.50 | 100.00 |
| 56 | 73.09 | 13.67 | 5.41 | 0.08 | 1.80 | 1.40 | 3.99 | 0.58 | 100.00 |
| 57 | 69.38 | 14.09 | 5.14 | 0.24 | 3.60 | 2.21 | 4.49 | 0.85 | 100.00 |
| 58 | 69.27 | 14.57 | 5.02 | 0.25 | 2.75 | 2.45 | 4.72 | 0.94 | 100.00 |
| 59 | 71.70 | 13.63 | 5.19 | 0.18 | 2.35 | 1.92 | 4.41 | 0.62 | 100.00 |
| 60 | 71.41 | 14.09 | 5.32 | 0.12 | 2.09 | 1.61 | 4.73 | 0.52 | 100.00 |
| 61 | 72.87 | 13.51 | 5.45 | 0.16 | 2.18 | 1.19 | 4.21 | 0.30 | 100.00 |
| 62 | 72.78 | 13.36 | 5.28 | 0.14 | 2.72 | 1.29 | 3.98 | 0.39 | 100.00 |
| 63 | 72.40 | 13.91 | 5.27 | 0.07 | 2.22 | 1.41 | 4.21 | 0.39 | 100.00 |
| 64 | 71.18 | 13.98 | 5.21 | 0.23 | 2.57 | 1.80 | 4.27 | 0.59 | 100.00 |
| 65 | 72.03 | 13.49 | 5.21 | 0.19 | 2.39 | 1.65 | 4.30 | 0.72 | 100.00 |
| 66 | 70.01 | 14.08 | 5.04 | 0.22 | 3.15 | 1.98 | 4.70 | 0.79 | 100.00 |
| 67 | 72.29 | 13.67 | 5.35 | 0.18 | 2.45 | 1.30 | 4.12 | 0.45 | 100.00 |
| 68 | 72.92 | 13.45 | 5.36 | 0.10 | 2.06 | 1.48 | 4.35 | 0.26 | 100.00 |
| 69 | 73.34 | 13.78 | 5.34 | 0.13 | 1.60 | 1.41 | 4.01 | 0.24 | 100.00 |
| 70 | 72.18 | 13.55 | 5.32 | 0.14 | 2.47 | 1.35 | 4.57 | 0.32 | 100.00 |
| 71 | 72.26 | 14.18 | 5.41 | 0.10 | 2.34 | 1.09 | 4.28 | 0.27 | 100.00 |
| 72 | 72.72 | 13.93 | 5.39 | 0.09 | 1.90 | 1.15 | 4.15 | 0.46 | 100.00 |
| 73 | 72.72 | 13.40 | 5.37 | 0.15 | 2.01 | 1.22 | 4.51 | 0.37 | 100.00 |
| 74 | 72.43 | 14.01 | 5.36 | 0.18 | 2.08 | 1.26 | 4.41 | 0.27 | 100.00 |
| 75 | 72.81 | 13.70 | 5.34 | 0.15 | 2.51 | 1.21 | 4.05 | 0.23 | 100.00 |
| 76 | 72.64 | 13.78 | 5.45 | 0.13 | 1.84 | 1.24 | 4.39 | 0.39 | 100.00 |
| 77 | 73.18 | 13.80 | 5.30 | 0.17 | 2.02 | 0.99 | 4.17 | 0.30 | 100.00 |
| 78 | 72.71 | 13.73 | 5.35 | 0.13 | 2.47 | 1.25 | 4.12 | 0.22 | 100.00 |
| 79 | 72.84 | 13.93 | 5.52 | 0.16 | 2.39 | 1.04 | 3.90 | 0.19 | 100.00 |
| Profile 3 | | | | | | | | | |
| 1 | 73.06 | 13.34 | 5.45 | 0.12 | 2.18 | 1.32 | 4.06 | 0.35 | 100.00 |
| 2 | 72.55 | 13.93 | 5.17 | 0.17 | 2.25 | 1.15 | 4.50 | 0.28 | 100.00 |
| 3 | 72.96 | 13.88 | 5.24 | 0.19 | 1.79 | 1.23 | 4.35 | 0.35 | 100.00 |
| 4 | 72.94 | 13.63 | 5.33 | 0.12 | 1.85 | 1.34 | 4.24 | 0.41 | 100.00 |
| 5 | 72.70 | 13.54 | 5.44 | 0.14 | 2.25 | 1.33 | 4.03 | 0.38 | 100.00 |
| 6 | 72.99 | 13.56 | 5.27 | 0.17 | 1.95 | 1.45 | 4.05 | 0.54 | 100.00 |
| 7 | 70.52 | 14.04 | 5.17 | 0.20 | 3.22 | 1.84 | 4.23 | 0.68 | 100.00 |
| 8 | 72.00 | 13.57 | 5.21 | 0.20 | 2.58 | 1.83 | 4.00 | 0.52 | 100.00 |
| 9 | 72.41 | 13.90 | 5.30 | 0.14 | 2.03 | 1.37 | 4.04 | 0.62 | 100.00 |
| 10 | 68.82 | 14.58 | 4.90 | 0.26 | 3.36 | 2.58 | 4.48 | 0.87 | 100.00 |

| | | | | | | | | |
|---|---|---|---|---|---|---|---|---|
| 11 | 66.11 | 15.36 | 4.84 | 0.34 | 3.73 | 3.49 | 4.64 | 1.36 | 100.00 |
| 12 | 64.86 | 15.04 | 4.71 | 0.35 | 5.28 | 3.42 | 4.74 | 1.56 | 100.00 |
| 13 | 66.37 | 14.75 | 4.70 | 0.40 | 4.43 | 3.43 | 4.35 | 1.49 | 100.00 |
| 14 | 66.06 | 15.01 | 4.66 | 0.29 | 4.13 | 3.29 | 4.92 | 1.56 | 100.00 |
| 15 | 68.07 | 14.30 | 4.82 | 0.29 | 3.66 | 2.96 | 4.37 | 1.43 | 100.00 |
| 16 | 68.20 | 14.03 | 4.84 | 0.34 | 3.73 | 3.08 | 4.29 | 1.42 | 100.00 |
| 17 | 66.43 | 14.43 | 4.54 | 0.40 | 4.25 | 3.40 | 4.89 | 1.53 | 100.00 |
| 18 | 65.20 | 15.15 | 4.53 | 0.40 | 4.15 | 3.75 | 4.46 | 1.93 | 100.00 |
| 19 | 64.79 | 15.21 | 4.38 | 0.43 | 4.10 | 4.05 | 4.67 | 2.06 | 100.00 |
| 20 | 64.57 | 14.88 | 4.40 | 0.44 | 4.78 | 4.26 | 4.62 | 2.00 | 100.00 |
| 21 | 63.58 | 15.19 | 4.36 | 0.40 | 5.34 | 4.23 | 4.65 | 2.07 | 100.00 |
| 22 | 63.11 | 15.29 | 4.13 | 0.40 | 5.92 | 4.68 | 4.12 | 2.06 | 100.00 |
| 23 | 62.45 | 15.30 | 4.11 | 0.37 | 5.64 | 4.97 | 4.59 | 2.54 | 100.00 |
| 24 | 62.25 | 15.12 | 4.00 | 0.52 | 5.87 | 5.00 | 4.56 | 2.43 | 100.00 |
| 25 | 63.17 | 15.28 | 4.28 | 0.48 | 5.33 | 4.63 | 4.31 | 2.31 | 100.00 |
| 26 | 64.90 | 14.73 | 4.49 | 0.34 | 4.37 | 4.30 | 4.60 | 2.10 | 100.00 |
| 27 | 64.85 | 14.94 | 4.31 | 0.43 | 4.72 | 3.87 | 4.61 | 1.87 | 100.00 |
| 28 | 63.47 | 15.36 | 4.30 | 0.49 | 5.04 | 4.67 | 4.58 | 1.98 | 100.00 |
| 29 | 65.58 | 14.72 | 4.59 | 0.38 | 4.39 | 3.81 | 4.48 | 1.74 | 100.00 |
| 30 | 68.56 | 14.45 | 4.92 | 0.24 | 3.47 | 2.46 | 4.63 | 1.12 | 100.00 |
| 31 | 72.05 | 14.01 | 5.26 | 0.14 | 2.20 | 1.58 | 4.21 | 0.52 | 100.00 |
| 32 | 72.88 | 13.56 | 5.54 | 0.10 | 1.76 | 1.48 | 4.11 | 0.42 | 100.00 |
| 33 | 72.94 | 13.88 | 5.24 | 0.16 | 1.80 | 1.17 | 4.30 | 0.26 | 100.00 |
| 34 | 72.25 | 14.01 | 5.14 | 0.16 | 2.48 | 1.36 | 4.16 | 0.34 | 100.00 |
| 35 | 70.97 | 14.14 | 5.26 | 0.16 | 2.67 | 1.62 | 4.46 | 0.60 | 100.00 |
| 36 | 69.33 | 14.25 | 4.89 | 0.28 | 3.46 | 2.28 | 4.39 | 1.03 | 100.00 |
| 37 | 69.75 | 14.46 | 4.83 | 0.26 | 2.63 | 2.81 | 4.22 | 1.02 | 100.00 |
| 38 | 69.83 | 14.38 | 5.25 | 0.13 | 3.07 | 2.12 | 4.58 | 0.58 | 100.00 |
| 39 | 70.01 | 14.38 | 5.20 | 0.17 | 3.17 | 1.87 | 4.52 | 0.67 | 100.00 |
| 40 | 72.70 | 13.77 | 5.15 | 0.12 | 2.06 | 1.41 | 4.05 | 0.45 | 100.00 |
| 41 | 72.73 | 13.88 | 5.44 | 0.12 | 1.98 | 1.18 | 4.15 | 0.29 | 100.00 |
| 42 | 73.53 | 13.63 | 5.43 | 0.17 | 1.29 | 1.36 | 4.24 | 0.35 | 100.00 |
| 43 | 72.91 | 13.54 | 5.36 | 0.12 | 1.94 | 1.43 | 4.15 | 0.30 | 100.00 |
| 44 | 72.08 | 13.80 | 5.37 | 0.13 | 2.35 | 1.54 | 4.14 | 0.51 | 100.00 |
| 45 | 70.39 | 13.88 | 5.23 | 0.23 | 2.98 | 1.71 | 4.65 | 0.73 | 100.00 |
| 46 | 71.14 | 13.98 | 5.36 | 0.18 | 2.44 | 1.81 | 4.43 | 0.48 | 100.00 |
| 47 | 72.83 | 13.65 | 5.63 | 0.12 | 1.85 | 1.41 | 3.93 | 0.53 | 100.00 |
| 48 | 72.52 | 13.63 | 5.42 | 0.19 | 1.91 | 1.55 | 4.11 | 0.49 | 100.00 |
| 49 | 72.83 | 13.53 | 5.38 | 0.12 | 1.63 | 1.60 | 4.38 | 0.51 | 100.00 |
| 50 | 69.71 | 14.74 | 5.23 | 0.22 | 3.18 | 1.88 | 4.41 | 0.63 | 100.00 |
| 51 | 71.72 | 13.91 | 5.36 | 0.08 | 2.16 | 1.54 | 4.49 | 0.46 | 100.00 |

| | | | | | | | | |
|---|---|---|---|---|---|---|---|---|
| 52 | 71.49 | 14.00 | 5.38 | 0.13 | 2.61 | 1.76 | 4.02 | 0.61 | 100.00 |
| 53 | 73.51 | 13.49 | 5.21 | 0.09 | 1.39 | 1.50 | 4.29 | 0.42 | 100.00 |
| 54 | 69.77 | 14.04 | 5.09 | 0.23 | 2.91 | 2.26 | 4.63 | 0.97 | 100.00 |
| 55 | 70.58 | 14.10 | 5.03 | 0.14 | 2.85 | 1.96 | 4.37 | 0.88 | 100.00 |
| 56 | 70.15 | 14.18 | 5.05 | 0.20 | 3.01 | 2.16 | 4.13 | 0.90 | 100.00 |
| 57 | 71.72 | 13.76 | 5.23 | 0.19 | 2.82 | 1.60 | 4.15 | 0.50 | 100.00 |
| 58 | 71.72 | 13.66 | 5.08 | 0.22 | 2.54 | 1.83 | 4.33 | 0.61 | 100.00 |
| 59 | 70.85 | 14.37 | 5.15 | 0.21 | 2.64 | 1.96 | 3.99 | 0.68 | 100.00 |
| 60 | 70.42 | 14.05 | 5.14 | 0.23 | 2.97 | 2.14 | 4.24 | 0.81 | 100.00 |
| 61 | 71.46 | 13.28 | 5.33 | 0.19 | 2.90 | 1.85 | 4.19 | 0.79 | 100.00 |
| 62 | 69.93 | 14.34 | 5.07 | 0.18 | 2.80 | 2.28 | 4.29 | 0.92 | 100.00 |
| 63 | 71.40 | 13.59 | 5.11 | 0.14 | 2.97 | 1.81 | 4.13 | 0.75 | 100.00 |
| 64 | 70.73 | 13.83 | 5.12 | 0.20 | 2.64 | 2.11 | 4.42 | 0.95 | 100.00 |
| 65 | 70.17 | 14.27 | 5.22 | 0.22 | 2.80 | 2.02 | 4.48 | 0.82 | 100.00 |
| 66 | 69.85 | 13.72 | 4.90 | 0.20 | 3.43 | 2.32 | 4.73 | 0.84 | 100.00 |
| 67 | 68.74 | 14.69 | 5.03 | 0.22 | 3.22 | 2.35 | 4.83 | 0.92 | 100.00 |
| 68 | 71.90 | 13.89 | 5.21 | 0.14 | 2.21 | 1.75 | 4.27 | 0.56 | 100.00 |
| 69 | 72.39 | 14.00 | 5.28 | 0.16 | 1.98 | 1.33 | 4.32 | 0.53 | 100.00 |
| 70 | 72.82 | 13.47 | 5.39 | 0.14 | 2.18 | 1.25 | 4.12 | 0.57 | 100.00 |
| 71 | 70.98 | 14.09 | 5.31 | 0.16 | 2.97 | 1.58 | 4.35 | 0.56 | 100.00 |
| 72 | 72.47 | 13.51 | 5.30 | 0.12 | 2.54 | 1.57 | 4.02 | 0.47 | 100.00 |
| 73 | 72.46 | 13.73 | 5.35 | 0.16 | 2.36 | 1.54 | 3.89 | 0.45 | 100.00 |
| 74 | 72.65 | 14.12 | 5.31 | 0.13 | 1.56 | 1.28 | 4.32 | 0.47 | 100.00 |
| 75 | 71.90 | 14.09 | 5.30 | 0.18 | 2.22 | 1.77 | 3.99 | 0.55 | 100.00 |
| 76 | 71.31 | 14.03 | 5.22 | 0.12 | 2.52 | 1.88 | 4.26 | 0.65 | 100.00 |
| 77 | 71.72 | 13.83 | 5.19 | 0.18 | 2.37 | 1.66 | 4.30 | 0.60 | 100.00 |
| 78 | 69.44 | 14.21 | 5.06 | 0.16 | 3.83 | 1.91 | 4.67 | 0.72 | 100.00 |
| 79 | 69.65 | 14.39 | 5.24 | 0.24 | 3.28 | 2.04 | 4.33 | 0.78 | 100.00 |
| 80 | 72.45 | 13.61 | 5.47 | 0.10 | 1.83 | 1.63 | 4.26 | 0.59 | 100.00 |
| 81 | 72.03 | 14.47 | 5.36 | 0.14 | 1.80 | 1.36 | 4.42 | 0.42 | 100.00 |
| 82 | 72.72 | 13.79 | 5.20 | 0.16 | 2.27 | 1.44 | 4.13 | 0.29 | 100.00 |
| 83 | 73.04 | 13.46 | 5.34 | 0.18 | 1.51 | 1.44 | 4.55 | 0.46 | 100.00 |
| 84 | 70.86 | 14.23 | 5.27 | 0.15 | 2.63 | 1.71 | 4.50 | 0.60 | 100.00 |
| 85 | 72.68 | 14.00 | 5.20 | 0.13 | 1.82 | 1.38 | 4.18 | 0.52 | 100.00 |
| 86 | 70.36 | 13.66 | 5.11 | 0.21 | 3.41 | 1.84 | 4.61 | 0.52 | 100.00 |
| 87 | 70.31 | 13.81 | 5.02 | 0.21 | 2.69 | 2.32 | 4.67 | 0.85 | 100.00 |
| 88 | 71.18 | 13.63 | 5.13 | 0.19 | 2.94 | 1.88 | 4.19 | 0.75 | 100.00 |
| 89 | 70.26 | 13.96 | 5.29 | 0.21 | 2.77 | 2.01 | 4.70 | 0.80 | 100.00 |
| 90 | 72.07 | 13.95 | 5.26 | 0.13 | 2.37 | 1.41 | 4.06 | 0.59 | 100.00 |
| 91 | 70.85 | 13.91 | 5.33 | 0.18 | 2.65 | 2.15 | 4.00 | 0.77 | 100.00 |
| 92 | 70.51 | 13.48 | 5.10 | 0.20 | 3.35 | 2.03 | 4.22 | 1.00 | 100.00 |

| | | | | | | | | |
|---|---|---|---|---|---|---|---|---|
| 93 | 68.21 | 14.44 | 4.87 | 0.32 | 3.29 | 2.80 | 4.52 | 1.22 | 100.00 |
| 94 | 66.47 | 14.82 | 4.72 | 0.29 | 3.96 | 3.60 | 4.68 | 1.46 | 100.00 |
| 95 | 66.43 | 15.07 | 4.75 | 0.33 | 3.80 | 3.32 | 4.62 | 1.49 | 100.00 |
| 96 | 70.62 | 13.69 | 5.05 | 0.19 | 3.18 | 1.99 | 4.47 | 0.81 | 100.00 |
| 97 | 71.66 | 14.16 | 5.42 | 0.15 | 2.01 | 1.64 | 4.23 | 0.64 | 100.00 |
| 98 | 72.22 | 13.55 | 5.52 | 0.10 | 1.87 | 1.48 | 4.36 | 0.52 | 100.00 |
| 99 | 71.58 | 13.80 | 5.30 | 0.24 | 2.52 | 1.56 | 4.48 | 0.52 | 100.00 |
| 100 | 72.95 | 13.78 | 5.50 | 0.13 | 2.01 | 1.00 | 4.12 | 0.30 | 100.00 |
| 101 | 72.13 | 13.79 | 5.46 | 0.15 | 2.27 | 1.27 | 4.42 | 0.30 | 100.00 |
| 102 | 72.74 | 13.61 | 5.31 | 0.14 | 2.38 | 1.13 | 4.09 | 0.39 | 100.00 |
| 103 | 72.91 | 13.41 | 5.28 | 0.09 | 2.18 | 1.24 | 4.33 | 0.42 | 100.00 |
| 104 | 72.65 | 13.85 | 5.35 | 0.16 | 2.06 | 1.27 | 4.08 | 0.49 | 100.00 |
| 105 | 72.83 | 13.52 | 5.36 | 0.17 | 2.16 | 1.30 | 4.14 | 0.46 | 100.00 |
| 106 | 71.15 | 14.16 | 5.16 | 0.16 | 2.70 | 1.76 | 4.31 | 0.50 | 100.00 |
| 107 | 71.82 | 14.02 | 5.53 | 0.13 | 2.25 | 1.30 | 4.48 | 0.45 | 100.00 |
| 108 | 72.01 | 14.09 | 5.32 | 0.18 | 2.15 | 1.45 | 4.35 | 0.45 | 100.00 |
| 109 | 71.80 | 13.85 | 5.35 | 0.17 | 2.40 | 1.61 | 4.42 | 0.41 | 100.00 |
| 110 | 72.25 | 13.78 | 5.35 | 0.19 | 1.93 | 1.67 | 4.39 | 0.41 | 100.00 |
| 111 | 72.47 | 14.11 | 5.32 | 0.10 | 2.02 | 1.24 | 4.28 | 0.33 | 100.00 |
| 112 | 72.92 | 13.72 | 5.35 | 0.16 | 2.35 | 1.07 | 4.09 | 0.34 | 100.00 |
| 113 | 73.07 | 13.55 | 5.39 | 0.21 | 1.75 | 1.18 | 4.40 | 0.39 | 100.00 |
| 114 | 72.17 | 14.12 | 5.47 | 0.15 | 2.23 | 1.12 | 4.33 | 0.29 | 100.00 |
| 115 | 72.12 | 13.66 | 5.50 | 0.13 | 2.60 | 1.30 | 4.39 | 0.30 | 100.00 |
| 116 | 71.81 | 14.22 | 5.29 | 0.18 | 2.42 | 1.40 | 4.23 | 0.36 | 100.00 |
| 117 | 72.49 | 13.61 | 5.39 | 0.12 | 2.43 | 1.23 | 4.37 | 0.30 | 100.00 |
| 118 | 73.16 | 13.85 | 5.43 | 0.11 | 1.91 | 1.14 | 4.02 | 0.31 | 100.00 |
| **Profile 4** | | | | | | | | | |
| 1 | 72.05 | 13.86 | 5.38 | 0.12 | 2.15 | 1.30 | 4.67 | 0.47 | 100.00 |
| 2 | 72.44 | 13.12 | 5.35 | 0.19 | 2.56 | 1.42 | 4.40 | 0.53 | 100.00 |
| 3 | 71.75 | 13.53 | 5.56 | 0.15 | 2.35 | 1.53 | 4.31 | 0.64 | 100.00 |
| 4 | 71.65 | 13.66 | 5.27 | 0.20 | 2.59 | 1.41 | 4.55 | 0.63 | 100.00 |
| 5 | 71.19 | 13.23 | 5.47 | 0.16 | 3.05 | 1.45 | 4.64 | 0.68 | 100.00 |
| 6 | 71.07 | 13.35 | 5.35 | 0.16 | 2.71 | 1.82 | 4.46 | 0.69 | 100.00 |
| 7 | 70.57 | 14.21 | 5.09 | 0.13 | 3.02 | 1.87 | 4.50 | 0.62 | 100.00 |
| 8 | 70.23 | 14.36 | 5.11 | 0.21 | 2.60 | 2.17 | 4.67 | 0.62 | 100.00 |
| 9 | 70.08 | 14.31 | 5.20 | 0.17 | 2.83 | 1.91 | 4.66 | 0.80 | 100.00 |
| 10 | 69.03 | 14.47 | 5.11 | 0.29 | 3.31 | 2.07 | 4.82 | 0.85 | 100.00 |
| 11 | 69.00 | 14.96 | 5.20 | 0.28 | 2.86 | 2.06 | 4.70 | 0.70 | 100.00 |
| 12 | 68.71 | 14.47 | 5.10 | 0.28 | 3.03 | 2.22 | 5.05 | 0.93 | 100.00 |
| 13 | 68.63 | 14.85 | 5.17 | 0.19 | 3.29 | 2.20 | 4.54 | 1.00 | 100.00 |
| 14 | 69.25 | 13.89 | 5.23 | 0.21 | 3.41 | 2.08 | 4.74 | 0.97 | 100.00 |

| | | | | | | | | |
|---|---|---|---|---|---|---|---|---|
| 15 | 70.60 | 13.52 | 5.24 | 0.24 | 2.88 | 2.18 | 4.46 | 0.88 | 100.00 |
| 16 | 71.19 | 13.92 | 5.14 | 0.16 | 2.27 | 1.99 | 4.58 | 0.76 | 100.00 |
| 17 | 70.74 | 13.77 | 5.23 | 0.15 | 2.89 | 1.99 | 4.34 | 0.86 | 100.00 |
| 18 | 71.01 | 14.05 | 5.30 | 0.17 | 2.28 | 1.75 | 4.73 | 0.71 | 100.00 |
| 19 | 70.93 | 14.25 | 5.26 | 0.17 | 2.38 | 1.61 | 4.49 | 0.85 | 100.00 |
| 20 | 69.95 | 14.16 | 5.17 | 0.22 | 2.75 | 2.06 | 4.78 | 0.91 | 100.00 |
| 21 | 69.68 | 14.79 | 5.26 | 0.20 | 2.44 | 2.12 | 4.52 | 0.85 | 100.00 |
| 22 | 69.80 | 13.91 | 5.35 | 0.15 | 3.46 | 1.77 | 4.80 | 0.70 | 100.00 |
| 23 | 70.43 | 13.97 | 5.15 | 0.22 | 2.98 | 1.98 | 4.44 | 0.70 | 100.00 |
| 24 | 68.57 | 14.29 | 5.17 | 0.22 | 3.21 | 2.45 | 4.78 | 1.01 | 100.00 |
| 25 | 67.32 | 14.74 | 5.05 | 0.25 | 3.84 | 2.59 | 5.02 | 1.05 | 100.00 |
| 26 | 69.19 | 14.68 | 4.95 | 0.16 | 2.32 | 2.50 | 5.00 | 1.04 | 100.00 |
| 27 | 70.33 | 13.66 | 5.10 | 0.22 | 3.28 | 1.98 | 4.49 | 0.90 | 100.00 |
| 28 | 69.25 | 14.36 | 5.02 | 0.17 | 3.11 | 2.26 | 4.54 | 0.95 | 100.00 |
| 29 | 69.05 | 14.36 | 5.19 | 0.24 | 3.51 | 2.18 | 4.49 | 0.78 | 100.00 |
| 30 | 70.83 | 13.79 | 5.21 | 0.20 | 2.62 | 1.96 | 4.49 | 0.88 | 100.00 |
| 31 | 70.43 | 14.29 | 5.14 | 0.23 | 2.63 | 2.00 | 4.43 | 0.78 | 100.00 |
| 32 | 69.42 | 14.35 | 5.14 | 0.26 | 3.61 | 1.90 | 4.56 | 0.73 | 100.00 |
| 33 | 70.32 | 14.12 | 5.15 | 0.21 | 2.69 | 2.18 | 4.47 | 0.83 | 100.00 |
| 34 | 70.60 | 14.23 | 5.30 | 0.22 | 2.92 | 1.68 | 4.28 | 0.65 | 100.00 |
| 35 | 71.44 | 13.67 | 5.32 | 0.15 | 2.13 | 1.92 | 4.46 | 0.69 | 100.00 |
| 36 | 71.88 | 13.80 | 5.34 | 0.13 | 2.36 | 1.54 | 4.32 | 0.53 | 100.00 |
| 37 | 72.42 | 13.66 | 5.41 | 0.13 | 2.01 | 1.36 | 4.25 | 0.51 | 100.00 |
| 38 | 72.49 | 13.86 | 5.35 | 0.15 | 1.91 | 1.27 | 4.44 | 0.49 | 100.00 |
| 39 | 72.73 | 14.01 | 5.36 | 0.10 | 2.09 | 1.13 | 4.02 | 0.31 | 100.00 |
| 40 | 73.64 | 13.50 | 5.50 | 0.14 | 1.68 | 1.28 | 4.00 | 0.25 | 100.00 |
| 41 | 72.74 | 13.63 | 5.53 | 0.10 | 1.92 | 1.27 | 4.54 | 0.27 | 100.00 |
| 42 | 72.77 | 13.79 | 5.44 | 0.09 | 1.86 | 1.44 | 4.17 | 0.26 | 100.00 |
| 43 | 73.07 | 13.47 | 5.39 | 0.12 | 1.87 | 1.16 | 4.42 | 0.51 | 100.00 |
| 44 | 72.38 | 13.66 | 5.32 | 0.12 | 2.08 | 1.54 | 4.11 | 0.55 | 100.00 |
| 45 | 71.21 | 13.70 | 5.11 | 0.12 | 3.07 | 1.72 | 4.19 | 0.83 | 100.00 |
| 46 | 67.86 | 14.34 | 4.98 | 0.22 | 3.33 | 3.09 | 4.88 | 1.22 | 100.00 |
| 47 | 66.53 | 15.12 | 4.86 | 0.32 | 3.94 | 3.07 | 4.61 | 1.49 | 100.00 |
| 48 | 67.34 | 15.00 | 4.83 | 0.31 | 3.31 | 2.96 | 4.79 | 1.47 | 100.00 |
| 49 | 67.18 | 14.69 | 4.97 | 0.33 | 3.85 | 2.87 | 4.77 | 1.34 | 100.00 |
| 50 | 67.69 | 14.52 | 5.12 | 0.29 | 3.55 | 2.72 | 4.75 | 1.28 | 100.00 |
| 51 | 68.95 | 13.97 | 4.89 | 0.20 | 3.09 | 2.65 | 4.73 | 1.28 | 100.00 |
| 52 | 68.99 | 13.86 | 5.07 | 0.23 | 3.21 | 2.56 | 4.85 | 1.11 | 100.00 |
| 53 | 68.29 | 13.93 | 5.06 | 0.18 | 3.95 | 2.49 | 4.65 | 1.20 | 100.00 |
| 54 | 67.99 | 13.87 | 4.91 | 0.26 | 4.15 | 2.89 | 4.63 | 1.30 | 100.00 |
| 55 | 67.59 | 14.35 | 4.92 | 0.24 | 3.79 | 3.12 | 4.59 | 1.34 | 100.00 |

| | | | | | | | | |
|---|---|---|---|---|---|---|---|---|
| 56 | 67.53 | 14.65 | 4.80 | 0.25 | 3.73 | 2.92 | 4.54 | 1.41 | 100.00 |
| 57 | 68.25 | 14.35 | 5.06 | 0.34 | 3.55 | 2.70 | 4.59 | 1.16 | 100.00 |
| 58 | 69.66 | 13.93 | 5.02 | 0.15 | 2.84 | 2.55 | 4.41 | 1.14 | 100.00 |
| 59 | 70.20 | 13.62 | 5.13 | 0.21 | 3.35 | 2.09 | 4.45 | 0.92 | 100.00 |
| 60 | 67.32 | 14.69 | 4.86 | 0.25 | 3.60 | 3.07 | 4.64 | 1.45 | 100.00 |
| 61 | 66.70 | 14.47 | 4.71 | 0.32 | 3.87 | 3.42 | 4.73 | 1.56 | 100.00 |
| 62 | 67.62 | 14.40 | 4.92 | 0.31 | 3.57 | 2.93 | 4.59 | 1.38 | 100.00 |
| 63 | 67.55 | 14.05 | 4.95 | 0.21 | 4.29 | 2.98 | 4.62 | 1.35 | 100.00 |
| 64 | 65.63 | 14.92 | 4.66 | 0.33 | 4.12 | 3.69 | 4.84 | 1.80 | 100.00 |
| 65 | 64.21 | 14.67 | 4.57 | 0.40 | 4.73 | 4.36 | 4.86 | 1.99 | 100.00 |
| 66 | 62.23 | 15.66 | 4.20 | 0.51 | 5.12 | 4.99 | 4.64 | 2.41 | 100.00 |
| 67 | 62.49 | 15.23 | 4.28 | 0.43 | 5.04 | 5.08 | 4.65 | 2.73 | 100.00 |
| 68 | 62.13 | 14.90 | 4.25 | 0.46 | 5.63 | 5.19 | 4.87 | 2.58 | 100.00 |
| 69 | 62.59 | 15.04 | 4.21 | 0.48 | 5.05 | 5.14 | 4.67 | 2.62 | 100.00 |
| 70 | 62.38 | 15.28 | 4.18 | 0.45 | 5.41 | 4.78 | 4.85 | 2.55 | 100.00 |
| 71 | 62.37 | 15.24 | 4.24 | 0.46 | 5.42 | 5.12 | 4.76 | 2.34 | 100.00 |
| 72 | 61.73 | 14.83 | 4.07 | 0.49 | 6.06 | 4.94 | 5.12 | 2.49 | 100.00 |
| 73 | 61.48 | 15.47 | 4.16 | 0.49 | 5.71 | 5.10 | 4.63 | 2.57 | 100.00 |
| 74 | 59.90 | 15.91 | 3.90 | 0.59 | 6.16 | 5.90 | 4.64 | 3.00 | 100.00 |
| 75 | 58.49 | 15.89 | 3.66 | 0.61 | 6.58 | 6.43 | 4.73 | 3.24 | 100.00 |
| 76 | 59.85 | 15.18 | 3.79 | 0.60 | 6.78 | 5.87 | 4.47 | 3.22 | 100.00 |
| 77 | 60.81 | 15.54 | 3.94 | 0.48 | 6.14 | 5.34 | 4.70 | 2.95 | 100.00 |
| 78 | 61.49 | 15.24 | 4.03 | 0.47 | 5.60 | 5.57 | 4.96 | 2.64 | 100.00 |
| 79 | 66.96 | 14.33 | 4.95 | 0.35 | 3.80 | 3.29 | 4.55 | 1.76 | 100.00 |
| 80 | 65.34 | 15.25 | 4.87 | 0.28 | 4.05 | 3.81 | 4.52 | 1.82 | 100.00 |
| 81 | 68.38 | 14.09 | 5.09 | 0.28 | 3.30 | 2.64 | 4.73 | 1.39 | 100.00 |
| 82 | 67.21 | 14.56 | 4.94 | 0.31 | 3.00 | 3.40 | 4.95 | 1.47 | 100.00 |
| 83 | 67.94 | 13.98 | 4.84 | 0.25 | 3.38 | 3.15 | 4.74 | 1.48 | 100.00 |
| 84 | 69.43 | 13.15 | 4.92 | 0.30 | 3.68 | 2.63 | 4.64 | 1.10 | 100.00 |
| 85 | 68.94 | 13.89 | 4.96 | 0.22 | 3.51 | 2.40 | 4.62 | 1.31 | 100.00 |
| 86 | 69.04 | 13.67 | 5.27 | 0.25 | 2.92 | 2.97 | 4.58 | 1.30 | 100.00 |
| 87 | 68.92 | 13.60 | 4.75 | 0.30 | 3.20 | 2.94 | 4.56 | 1.47 | 100.00 |
| 88 | 68.69 | 13.96 | 4.92 | 0.32 | 3.59 | 2.76 | 4.50 | 1.12 | 100.00 |
| 89 | 69.30 | 13.67 | 4.97 | 0.20 | 4.02 | 2.37 | 4.42 | 0.96 | 100.00 |
| 90 | 70.19 | 13.36 | 5.17 | 0.15 | 2.76 | 2.37 | 4.60 | 1.06 | 100.00 |
| 91 | 69.63 | 13.96 | 5.01 | 0.16 | 3.53 | 2.22 | 4.33 | 1.07 | 100.00 |
| 92 | 69.38 | 13.66 | 5.15 | 0.25 | 3.75 | 2.14 | 4.67 | 1.01 | 100.00 |
| 93 | 68.50 | 14.27 | 5.20 | 0.31 | 3.77 | 2.27 | 4.83 | 0.84 | 100.00 |
| 94 | 68.08 | 15.44 | 5.06 | 0.30 | 2.96 | 2.62 | 4.60 | 0.91 | 100.00 |
| 95 | 70.06 | 13.94 | 5.18 | 0.26 | 3.17 | 1.95 | 4.54 | 0.90 | 100.00 |
| 96 | 72.63 | 13.27 | 5.20 | 0.16 | 2.55 | 1.39 | 4.11 | 0.66 | 100.00 |
| 97 | 72.16 | 13.62 | 5.38 | 0.16 | 2.29 | 1.51 | 4.12 | 0.61 | 100.00 |
| 98 | 72.84 | 13.84 | 5.44 | 0.12 | 1.76 | 1.22 | 4.17 | 0.51 | 100.00 |
| 99 | 72.53 | 13.75 | 5.35 | 0.15 | 2.01 | 1.20 | 4.57 | 0.42 | 100.00 |
| **Profile 5** | | | | | | | | | |
| 1 | 73.04 | 13.35 | 5.69 | 0.11 | 2.02 | 1.14 | 4.36 | 0.29 | 100.00 |
| 2 | 72.62 | 14.16 | 5.50 | 0.07 | 1.65 | 1.09 | 4.62 | 0.29 | 100.00 |
| 3 | 73.32 | 13.61 | 5.41 | 0.12 | 1.85 | 1.16 | 4.12 | 0.34 | 100.00 |
| 4 | 72.94 | 13.50 | 5.53 | 0.16 | 1.97 | 1.16 | 4.24 | 0.43 | 100.00 |
| 5 | 72.74 | 13.30 | 5.53 | 0.13 | 2.21 | 1.26 | 4.35 | 0.35 | 100.00 |
| 6 | 73.28 | 13.63 | 5.56 | 0.16 | 1.32 | 1.21 | 4.28 | 0.32 | 100.00 |
| 7 | 72.91 | 13.73 | 5.57 | 0.14 | 1.45 | 1.37 | 4.37 | 0.44 | 100.00 |
| 8 | 72.30 | 13.77 | 5.33 | 0.14 | 2.02 | 1.50 | 4.33 | 0.57 | 100.00 |
| 9 | 71.35 | 13.48 | 5.44 | 0.18 | 2.74 | 1.66 | 4.35 | 0.79 | 100.00 |
| 10 | 70.21 | 14.33 | 5.26 | 0.20 | 2.61 | 1.96 | 4.53 | 0.90 | 100.00 |
| 11 | 69.17 | 14.62 | 5.08 | 0.21 | 2.76 | 2.23 | 4.85 | 0.98 | 100.00 |
| 12 | 67.99 | 14.61 | 5.12 | 0.21 | 3.94 | 2.23 | 4.76 | 1.13 | 100.00 |
| 13 | 68.76 | 14.31 | 5.10 | 0.22 | 3.69 | 2.22 | 4.74 | 0.96 | 100.00 |
| 14 | 68.57 | 14.91 | 5.14 | 0.26 | 3.15 | 2.44 | 4.62 | 0.77 | 100.00 |
| 15 | 71.22 | 14.02 | 5.05 | 0.14 | 2.06 | 1.94 | 4.55 | 0.97 | 100.00 |
| 16 | 73.25 | 12.96 | 5.32 | 0.17 | 1.66 | 1.63 | 4.35 | 0.60 | 100.00 |
| 17 | 72.83 | 13.63 | 5.27 | 0.15 | 1.78 | 1.56 | 4.17 | 0.57 | 100.00 |
| 18 | 72.54 | 13.42 | 5.29 | 0.14 | 2.09 | 1.61 | 4.23 | 0.68 | 100.00 |
| 19 | 72.74 | 13.24 | 5.29 | 0.09 | 1.92 | 1.58 | 4.19 | 0.80 | 100.00 |
| 20 | 71.85 | 13.58 | 5.02 | 0.14 | 2.41 | 1.67 | 4.14 | 0.91 | 100.00 |
| 21 | 70.90 | 13.20 | 5.30 | 0.16 | 3.04 | 2.06 | 4.23 | 1.05 | 100.00 |
| 22 | 68.19 | 13.58 | 4.97 | 0.24 | 3.99 | 2.74 | 4.86 | 1.36 | 100.00 |
| 23 | 67.04 | 14.37 | 4.96 | 0.35 | 3.92 | 3.06 | 4.83 | 1.40 | 100.00 |
| 24 | 65.73 | 14.83 | 4.60 | 0.31 | 4.73 | 3.37 | 4.67 | 1.69 | 100.00 |
| 25 | 66.87 | 14.36 | 4.80 | 0.32 | 4.06 | 2.90 | 4.94 | 1.43 | 100.00 |
| 26 | 69.54 | 13.90 | 5.18 | 0.23 | 3.28 | 2.22 | 4.44 | 1.06 | 100.00 |
| 27 | 69.76 | 13.56 | 5.07 | 0.16 | 3.37 | 2.27 | 4.79 | 0.83 | 100.00 |
| 28 | 72.56 | 13.67 | 5.31 | 0.10 | 1.97 | 1.40 | 4.10 | 0.71 | 100.00 |
| 29 | 71.70 | 14.14 | 5.34 | 0.09 | 2.15 | 1.47 | 4.18 | 0.62 | 100.00 |
| 30 | 71.87 | 13.92 | 5.37 | 0.11 | 1.93 | 1.73 | 4.24 | 0.66 | 100.00 |
| 31 | 71.32 | 14.39 | 5.32 | 0.13 | 2.19 | 1.41 | 4.71 | 0.52 | 100.00 |
| 32 | 72.92 | 13.71 | 5.47 | 0.18 | 1.91 | 0.90 | 4.49 | 0.41 | 100.00 |
| 33 | 72.44 | 13.67 | 5.42 | 0.13 | 2.42 | 1.13 | 4.43 | 0.36 | 100.00 |
| 34 | 72.16 | 13.71 | 5.33 | 0.16 | 2.60 | 1.27 | 4.46 | 0.31 | 100.00 |
| 35 | 71.78 | 13.88 | 5.41 | 0.19 | 2.54 | 1.49 | 4.29 | 0.42 | 100.00 |
| 36 | 72.13 | 13.46 | 5.41 | 0.18 | 2.16 | 1.54 | 4.51 | 0.53 | 100.00 |
| 37 | 72.27 | 13.61 | 5.45 | 0.16 | 2.37 | 1.22 | 4.47 | 0.46 | 100.00 |

| | | | | | | | | |
|---|---|---|---|---|---|---|---|---|
| 38 | 72.69 | 13.46 | 5.41 | 0.13 | 2.13 | 1.35 | 4.24 | 0.44 | 100.00 |
| 39 | 72.10 | 13.59 | 5.31 | 0.18 | 2.19 | 1.52 | 4.15 | 0.79 | 100.00 |
| 40 | 70.17 | 13.10 | 5.07 | 0.20 | 3.70 | 2.21 | 4.56 | 1.00 | 100.00 |
| 41 | 67.30 | 13.98 | 4.97 | 0.33 | 4.18 | 3.08 | 4.67 | 1.41 | 100.00 |
| 42 | 64.14 | 15.55 | 4.49 | 0.36 | 4.63 | 4.24 | 4.65 | 1.88 | 100.00 |
| 43 | 65.58 | 14.83 | 4.44 | 0.41 | 4.46 | 3.89 | 4.58 | 1.81 | 100.00 |
| 44 | 72.40 | 13.02 | 5.25 | 0.17 | 2.24 | 1.83 | 3.99 | 0.98 | 100.00 |
| 45 | 71.80 | 13.77 | 5.26 | 0.12 | 2.45 | 1.45 | 4.36 | 0.61 | 100.00 |
| 46 | 72.62 | 13.98 | 5.55 | 0.15 | 1.75 | 1.09 | 4.16 | 0.52 | 100.00 |
| 47 | 72.04 | 13.54 | 5.57 | 0.12 | 2.18 | 1.31 | 4.55 | 0.42 | 100.00 |
| 48 | 73.02 | 13.25 | 5.31 | 0.16 | 1.82 | 1.46 | 4.34 | 0.62 | 100.00 |
| 49 | 67.16 | 14.11 | 4.89 | 0.33 | 3.41 | 3.43 | 5.04 | 1.63 | 100.00 |
| 50 | 61.77 | 15.27 | 4.03 | 0.50 | 5.61 | 5.36 | 4.72 | 2.68 | 100.00 |
| 51 | 59.00 | 16.19 | 3.56 | 0.61 | 6.14 | 6.59 | 4.57 | 3.22 | 100.00 |
| 52 | 57.50 | 16.10 | 3.49 | 0.64 | 6.30 | 7.39 | 4.63 | 3.69 | 100.00 |
| 53 | 58.20 | 15.85 | 3.47 | 0.66 | 7.00 | 6.60 | 4.51 | 3.47 | 100.00 |
| 54 | 58.54 | 15.70 | 3.87 | 0.68 | 6.22 | 6.78 | 4.69 | 3.33 | 100.00 |
| 55 | 58.23 | 16.14 | 3.39 | 0.58 | 6.52 | 7.18 | 4.44 | 3.40 | 100.00 |
| 56 | 58.14 | 15.94 | 3.51 | 0.62 | 7.24 | 6.66 | 4.32 | 3.32 | 100.00 |
| 57 | 59.38 | 15.62 | 3.81 | 0.56 | 6.42 | 6.29 | 4.64 | 2.89 | 100.00 |
| 58 | 59.74 | 15.52 | 3.86 | 0.60 | 6.12 | 6.25 | 4.56 | 3.26 | 100.00 |
| 59 | 59.94 | 15.86 | 3.72 | 0.58 | 5.94 | 6.42 | 4.59 | 2.95 | 100.00 |
| 60 | 59.28 | 15.76 | 3.72 | 0.53 | 5.88 | 6.59 | 4.61 | 3.23 | 100.00 |
| 61 | 59.35 | 15.61 | 3.75 | 0.56 | 6.76 | 6.09 | 4.64 | 3.11 | 100.00 |
| 62 | 58.78 | 15.27 | 3.78 | 0.56 | 7.77 | 6.14 | 4.48 | 3.13 | 100.00 |
| 63 | 59.16 | 15.40 | 3.61 | 0.62 | 6.97 | 6.28 | 4.56 | 3.25 | 100.00 |
| 64 | 59.07 | 15.64 | 3.57 | 0.49 | 6.54 | 6.42 | 4.84 | 3.25 | 100.00 |
| 65 | 60.01 | 15.20 | 3.73 | 0.50 | 6.80 | 6.02 | 4.53 | 3.21 | 100.00 |
| 66 | 64.91 | 14.48 | 4.43 | 0.35 | 4.48 | 4.00 | 5.01 | 2.13 | 100.00 |
| 67 | 70.79 | 13.01 | 4.96 | 0.18 | 2.96 | 2.38 | 4.46 | 1.27 | 100.00 |
| 68 | 70.83 | 13.16 | 5.03 | 0.17 | 3.21 | 2.16 | 4.30 | 1.11 | 100.00 |
| 69 | 70.12 | 12.95 | 4.96 | 0.18 | 3.59 | 2.37 | 4.40 | 1.33 | 100.00 |
| 70 | 67.04 | 14.21 | 4.82 | 0.25 | 3.65 | 3.43 | 4.70 | 1.68 | 100.00 |
| 71 | 68.19 | 14.14 | 5.03 | 0.33 | 3.26 | 2.86 | 4.75 | 1.39 | 100.00 |
| 72 | 64.93 | 15.17 | 4.56 | 0.35 | 4.70 | 4.03 | 4.42 | 1.84 | 100.00 |
| 73 | 63.93 | 15.00 | 4.38 | 0.49 | 4.57 | 4.76 | 4.54 | 2.29 | 100.00 |
| 74 | 59.67 | 15.73 | 3.82 | 0.46 | 6.14 | 6.31 | 4.87 | 2.84 | 100.00 |
| 75 | 60.81 | 15.68 | 3.96 | 0.57 | 5.62 | 5.63 | 4.79 | 2.81 | 100.00 |
| 76 | 61.94 | 15.56 | 4.07 | 0.46 | 5.49 | 4.85 | 4.62 | 2.79 | 100.00 |
| 77 | 63.52 | 14.95 | 4.38 | 0.38 | 5.47 | 4.42 | 4.63 | 2.13 | 100.00 |
| 78 | 67.28 | 14.38 | 4.99 | 0.25 | 3.71 | 3.18 | 4.81 | 1.40 | 100.00 |
| 79 | 71.54 | 13.83 | 5.34 | 0.19 | 2.21 | 1.77 | 4.32 | 0.71 | 100.00 |
| 80 | 72.32 | 13.61 | 5.49 | 0.19 | 1.93 | 1.36 | 4.65 | 0.41 | 100.00 |
| 81 | 72.47 | 13.59 | 5.29 | 0.12 | 2.30 | 1.29 | 4.32 | 0.39 | 100.00 |
| 82 | 72.65 | 13.54 | 5.52 | 0.16 | 1.65 | 1.42 | 4.37 | 0.40 | 100.00 |
| 83 | 73.30 | 13.67 | 5.41 | 0.12 | 2.02 | 1.10 | 3.95 | 0.25 | 100.00 |
| 84 | 72.45 | 13.61 | 5.49 | 0.10 | 2.24 | 1.33 | 4.42 | 0.35 | 100.00 |
| 85 | 72.86 | 13.95 | 5.51 | 0.17 | 1.56 | 1.18 | 4.37 | 0.41 | 100.00 |
| 86 | 72.77 | 13.77 | 5.38 | 0.11 | 2.01 | 1.25 | 4.33 | 0.38 | 100.00 |
| 87 | 73.28 | 13.63 | 5.39 | 0.15 | 1.97 | 1.30 | 3.95 | 0.34 | 100.00 |
| 88 | 72.38 | 13.47 | 5.36 | 0.17 | 2.21 | 1.56 | 4.28 | 0.44 | 100.00 |
| 89 | 72.10 | 13.60 | 5.20 | 0.20 | 2.31 | 1.68 | 4.19 | 0.69 | 100.00 |
| 90 | 70.06 | 14.10 | 5.10 | 0.29 | 2.69 | 2.14 | 4.50 | 0.80 | 100.00 |
| 91 | 70.28 | 13.94 | 5.08 | 0.18 | 2.96 | 2.05 | 4.55 | 0.83 | 100.00 |
| 92 | 68.30 | 14.61 | 4.98 | 0.24 | 3.86 | 2.47 | 4.36 | 1.17 | 100.00 |
| 93 | 69.53 | 14.04 | 4.96 | 0.23 | 3.12 | 2.16 | 4.78 | 0.90 | 100.00 |
| 94 | 71.07 | 14.18 | 5.13 | 0.21 | 2.20 | 1.79 | 4.66 | 0.72 | 100.00 |
| 95 | 71.78 | 13.61 | 5.33 | 0.14 | 2.48 | 1.59 | 4.42 | 0.50 | 100.00 |
| 96 | 72.46 | 13.67 | 5.30 | 0.14 | 2.05 | 1.42 | 4.39 | 0.56 | 100.00 |
| 97 | 72.81 | 13.75 | 5.37 | 0.15 | 1.85 | 1.30 | 4.27 | 0.32 | 100.00 |
| 98 | 71.88 | 14.03 | 5.48 | 0.15 | 2.17 | 1.40 | 4.51 | 0.37 | 100.00 |
| 99 | 72.26 | 13.63 | 5.51 | 0.12 | 2.51 | 1.19 | 4.54 | 0.24 | 100.00 |
| **Profile 6** | | | | | | | | | |
| 1 | 72.38 | 13.91 | 5.45 | 0.18 | 2.14 | 1.28 | 4.23 | 0.38 | 100.00 |
| 2 | 72.95 | 13.37 | 5.50 | 0.11 | 1.69 | 1.25 | 4.74 | 0.40 | 100.00 |
| 3 | 73.10 | 13.98 | 5.45 | 0.09 | 1.57 | 1.13 | 4.33 | 0.35 | 100.00 |
| 4 | 73.12 | 13.31 | 5.41 | 0.13 | 2.07 | 1.23 | 4.29 | 0.44 | 100.00 |
| 5 | 72.90 | 13.39 | 5.39 | 0.16 | 1.56 | 1.35 | 4.40 | 0.54 | 100.00 |
| 6 | 72.57 | 13.70 | 5.38 | 0.11 | 2.12 | 1.38 | 4.13 | 0.60 | 100.00 |
| 7 | 72.02 | 13.75 | 5.42 | 0.11 | 1.91 | 1.61 | 4.36 | 0.58 | 100.00 |
| 8 | 71.74 | 13.32 | 5.28 | 0.16 | 2.64 | 1.64 | 4.27 | 0.64 | 100.00 |
| 9 | 71.73 | 13.41 | 5.35 | 0.13 | 2.76 | 1.56 | 4.29 | 0.72 | 100.00 |
| 10 | 71.78 | 13.47 | 5.26 | 0.20 | 2.55 | 1.54 | 4.31 | 0.79 | 100.00 |
| 11 | 70.51 | 13.63 | 5.06 | 0.25 | 2.87 | 2.27 | 4.53 | 0.86 | 100.00 |
| 12 | 68.91 | 14.31 | 5.01 | 0.26 | 3.14 | 2.38 | 4.75 | 1.15 | 100.00 |
| 13 | 69.06 | 14.11 | 5.27 | 0.26 | 3.18 | 2.47 | 4.55 | 1.10 | 100.00 |
| 14 | 68.98 | 13.90 | 5.25 | 0.31 | 3.25 | 2.75 | 4.37 | 1.01 | 100.00 |
| 15 | 67.62 | 14.52 | 5.09 | 0.29 | 3.44 | 2.44 | 5.26 | 1.05 | 100.00 |
| 16 | 68.18 | 14.67 | 5.30 | 0.20 | 2.92 | 2.57 | 5.14 | 1.03 | 100.00 |
| 17 | 69.82 | 14.11 | 5.19 | 0.29 | 2.66 | 2.25 | 4.35 | 1.08 | 100.00 |
| 18 | 69.72 | 14.14 | 5.15 | 0.19 | 2.68 | 2.38 | 4.49 | 1.10 | 100.00 |